\documentclass[referee]{raa}           
\usepackage{graphics}
\usepackage{natbib}
\usepackage{amssymb,amsmath}
\usepackage{newtxtext,newtxmath}
\usepackage{booktabs,chemformula}
\usepackage{natbib}
\usepackage[utf8]{inputenc}
\usepackage{graphicx}
\usepackage{xurl}
\usepackage{lipsum}
\usepackage{booktabs}
\usepackage{seqsplit}
\usepackage{float}
\usepackage{array}
\usepackage{subcaption}
\usepackage[export]{adjustbox}
\usepackage{wrapfig}
\usepackage[para]{threeparttable}
\usepackage{caption}
\usepackage{multirow}
\usepackage{caption,stackengine}
\usepackage[T1]{fontenc}
\usepackage{ae,aecompl,arydshln}

\usepackage{amssymb}
\usepackage{lipsum}
\usepackage{xcolor}
\usepackage[colorlinks=true, linkcolor=blue, urlcolor=blue, citecolor=blue, pagebackref=true]{hyperref}
\usepackage{amsmath}
\usepackage{multirow}
\usepackage{hyperref}

\usepackage{multirow}
\newcommand{\asat}{\textit{AstroSat}}
\newcommand{\src}{{GRS 1915+105}}

\bibpunct{(}{)}{;}{a}{}{,}

\begin{document}


\title{Spectro-timing origin of large amplitude X-ray variability in GRS 1915+105 using AstroSat/LAXPC and SXT}

 \volnopage{ {\bf 20XX} Vol.\ {\bf X} No. {\bf XX}, 000--000}
   \setcounter{page}{1}

   \author{Shree Suman\inst{1}, Shuvajit Khatua\inst{1,2}, Vishal Jadoliya\inst{1}, Prathamesh Narayan Gupta\inst{3}, Mayukh Pahari\inst{1}
   }

   \institute{$^1$Department of Physics, Indian Institute of Technology, Hyderabad 502284, India, \\ 
              $^2$Department of Astronomy \& Astrophysics, Raman Research Institute, Bangalore\\
              $^3$Delhi Technological University, Shahbad Daulatpur Village, Rohini, New Delhi 110042\\
    Corresponding author email: mayukh@phy.iith.ac.in\\
}

\abstract{The origin of the large-amplitude, quasi-periodic X-ray flux variations in several classes of the Galactic microquasar GRS~1915+105 remains unresolved. We address this issue through flux-resolved, broadband (0.8-20 keV) spectral modelling and simultaneous covariance spectral analysis during two $\kappa$ and two $\omega$ class observations using \textit{AstroSat}/SXT and LAXPC. The lightcurves show strong, quasi-periodic oscillations involving rapid transitions between bright bursts and deep dips on timescales of a few tens of seconds. Flux-resolved spectroscopy indicates that high-flux intervals in both classes are dominated by a hot, optically thick accretion disc with steep Comptonized emission, whereas low-flux intervals correspond to a cooler or partially recessed disc and a harder coronal continuum. These transitions involve a systematic 1-2 keV drop in disc temperature and a pronounced hardening of the Comptonized component, with flux reductions of up to a factor of five. Using covariance spectra across 0.015-5 Hz, we show that the rapid coherent variability arises almost entirely from the disc, which exhibits strong energy-dependent variations, while the Comptonized component contributes minimally. The combined results suggest that radiation-pressure-driven structural changes in the disc, with a slower coronal response, produce the observed oscillations, consistent with cyclic disc evacuation and refilling in the $\kappa$ and $\omega$ classes.
}
\authorrunning{S. Suman et al. }            
\titlerunning{Signature of spectral variation in GRS 1915+105}  
   \maketitle

%
\section{Introduction}           
\label{sect:intro}

GRS 1915+105 is one of the most extraordinary Galactic X-ray binaries, discovered as a bright and highly variable source by \textit{Granat} in 1992 \citep{CastroTirado1992}. With a black hole mass of $\sim12.4\,M_{\odot}$ and a long orbital period of 33.5 days \citep{Steeghs2013, Reid2014}, the system hosts a rapidly spinning black hole whose extreme accretion rate often exceeds the Eddington limit. One of its most remarkable features is the presence of more than a dozen distinct X-ray variability classes \citep{Belloni2000}, including large-amplitude limit-cycle oscillations driven by accretion-disk instabilities \citep{Neilsen2011}. GRS\,1915+105 has also revealed powerful disc-jet coupling, with radio flares \cite{KleinWolt2002} and steady jets closely linked to transitions in its X-ray spectral states \citep{Fender2004}. Together, these discoveries have positioned GRS\,1915+105 as a cornerstone for studying accretion physics and radiation-pressure dominated disc instabilities in strong-gravity environments.

GRS 1915+105 is known for its dramatic flux and spectral variability, often undergoing rapid, quasi-periodic transitions between low-luminosity, spectrally hard states and bright, soft flare behaviour interpreted as limit–cycle oscillations driven by an unstable accretion flow \citep{Taam1984,Janiuk2000}. In detailed, phase-resolved spectro-timing studies of the $\rho$--state \citep{Neilsen2011}, the high amplitude oscillations are consistent with a thermal viscous, radiation pressure-driven instability in the inner disc that causes repeated evaporation and refilling of the innermost accretion flow, coupled with large changes in disc-wind properties and even hints of jet ejection triggered near minimum luminosity \citep{Zoghbi2016}. More recent theoretical work \citep{Janiuk2000b}, for instance, incorporating the effects of magnetic field strength \citep{ratheesh2021}and corona formation - has shown that modified radiation-pressure instability models can reproduce the observed amplitudes (factors of a few to $\sim$10) \citep{Rawat_2019}, periods (from $\sim$40\,s to $\sim$1500\,s) \citep{Nayakshin2000}, and in some cases even predict the suppression of limit cycles under particular accretion conditions.
Using \asat{} and \textit{NICER} data, \citet{Dhaka2024} showed that energy-dependent fractional rms during high and low fluxes cannot be explained using only varying coronal parameters.
Using 3-20 keV \textit{RXTE}/PCA data, \citep{Pahari2013} used flux-resolved spectroscopy of large-amplitude, quasi-regular oscillations (in particular those similar to the $\kappa$ and $\lambda$ classes) and showed that the `peak-minus-dip’ or difference spectra can be remarkably well described by a single $p$-free disc blackbody, suggesting that the flux variation arises largely from the appearance and disappearance of an additional accretion–disc component, rather than drastic changes in the Comptonizing corona. They argue that this behaviour lends strong empirical support to the hypothesis that the oscillations reflect a radiation-pressure-dominated inner disc undergoing limit-cycle instability, and that the true intrinsic disc variation may be much larger (potentially orders of magnitude) than the factor of ten seen in observed peak-versus-dip luminosities.

In addition to the $\rho$- and $\kappa$-type variability, the $\omega$ class has emerged as an important intermediate oscillatory state. Studies of long–term class evolution \citep{Pahari2010} show that during the $\omega$ class, the source exhibits nearly periodic fluctuations between higher and lower intensities, often with high–intensity plateaus several times brighter than the low–intensity intervals. These changes are accompanied by variations in spectral hardness, suggestive of oscillatory transitions between a disc–dominated soft state and a harder, Comptonization–dominated state. As the $\omega$ pattern evolves, the lower–flux dip intervals shorten, and the source eventually settles into a stable high–soft state, indicating changes in accretion rate and/or the disc-corona geometry. More recently, high–frequency QPOs in the $\kappa$ and $\omega$ classes \citep{Belloni2019} have further linked these states to rapid inner-flow dynamics, underlining their importance for understanding accretion-driven variability beyond the heartbeat regime.

The physical interpretation of the large-amplitude, quasi-periodic X-ray variability in GRS~1915+105 was first systematically established by \citet{Belloni1997}, who proposed that the recurrent transitions between bright, disc-dominated states and low-flux, spectrally harder intervals are driven by cyclic evacuation and refilling of the inner accretion disc due to a radiation-pressure instability. In this framework, the temporary disappearance or recession of the innermost disc during low-flux intervals leads to a reduced soft-photon supply and a harder Comptonized spectrum, while the subsequent refilling of the disc produces luminous soft states. This seminal model provided the foundation for the interpretation of the limit-cycle oscillations observed in several variability classes of GRS~1915+105.
 
In the present work, we explicitly build upon the \citet{Belloni1997} framework by performing flux-resolved, broadband (0.8--20~keV) spectral modelling together with covariance spectral analysis of the $\kappa$ and $\omega$ classes using AstroSat/SXT and LAXPC. Our approach allows us to quantitatively track the evolution of disc and coronal parameters across individual oscillation cycles and to directly identify the spectral component responsible for the rapid coherent variability.
Nevertheless, no single model yet accounts for the full phenomenology of GRS 1915+105: including the multiplicity of variability classes \citep{Bachetti2022}, their transitions, and the coupling between winds, jets, and the inner accretion disc ensuring that this extraordinary system remains a key laboratory for studying high accretion rate black hole physics.

In recent years, \asat{} observations have enabled detailed studies of variability classes and high-frequency quasi-periodic oscillations (HFQPOs) in GRS~1915+105, providing new insights into the dynamics of the inner accretion flow. Using AstroSat/LAXPC data, \citet{Majumder2022} investigated HFQPO properties across multiple variability classes and argued for a close connection between HFQPOs and changes in the innermost accretion geometry. More recently, \citet{Dhaka2025} performed a comprehensive timing and spectral study of flaring classes, demonstrating that variations in HFQPO frequency and strength are closely linked to changes in the inner disc and coronal properties. Similarly, \citet{Harikesh2025} examined the evolution of HFQPOs across different variability classes and suggested that these oscillations probe the dynamical state of the inner flow close to the black hole.
 
While these studies primarily focus on HFQPO phenomenology and its relation to inner-flow dynamics, the present work addresses a complementary aspect of the variability problem by concentrating on the origin of the large-amplitude, quasi-periodic flux oscillations themselves. By combining flux-resolved broadband spectroscopy with covariance spectral analysis, we directly identify the accretion disc as the dominant driver of the rapid coherent variability in the $\kappa$ and $\omega$ classes, thereby providing an essential physical context for the inner-flow changes inferred from HFQPO studies.
In this work, we conducted the first flux-resolved, broadband (0.8–20 keV) spectral and covariance-spectral analysis of two $\kappa$ and two $\omega$ class observations using \textit{AstroSat}/SXT and LAXPC. Flux-resolved spectroscopy shows that high-flux intervals in both classes are dominated by a hot, optically thick disc plus steep Comptonized emission, whereas low-flux intervals correspond to a cooler or partially recessed disc and a harder coronal continuum. Such transitions involve a 1–2 keV drop in disc temperature, pronounced hardening of the Comptonized component, and flux decreases of up to a factor of five. Using covariance spectra over 0.015–5 Hz, we find that the rapid coherent variability originates almost entirely from the disc, which shows strong energy-dependent fluctuations, while the Comptonized component contributes minimally. Together, these results indicate rapid, radiation-pressure-driven structural changes in the disc, with the corona responding more slowly and less coherently. The combined evidence supports a picture in which the $\kappa$ and $\omega$ classes are limit-cycle episodes of disc evacuation and refilling that produce the observed quasi-oscillations. 

The paper is organised as follows: In Section \ref{sec2}, we describe the X-ray observation and the corresponding data reduction method. Meanwhile, temporal analysis, and spectral analysis, along with their results, are provided in Section \ref{sec3} and \ref{sec4}, respectively. Covariance analysis and corresponding results are provided in Section \ref{sec5}. We discuss implications of the results and make conclusions in Section \ref{sec6}.

\section{OBSERVATION AND DATA REDUCTION}\label{sec2}
The data has been obtained through India’s first broad-band X-ray observatory, AstroSat, using simultaneous observations from both soft
X-ray telescope (SXT) and large area X-ray proportional counter (LAXPC) instruments \citep{singh2014}. SXT and LAXPC observation details of GRS 1915+105 are provided in Table \ref{tab1}.
\subsection{SXT}
The Soft X-ray Telescope (SXT) onboard \asat{} is a focusing X-ray instrument equipped with a CCD detector, providing imaging capabilities in the 0.3-7.0 keV energy range with medium spectral resolution \citep{Singh2017}. For GRS 1915+105, we analysed data from orbits that match those used for the LAXPC observations. The SXT data reduction was carried out using the \textsc{SXTPIPELINE v1.4b}\footnote{\url{http://www.tifr.res.in/~astrosat_sxt/sxtpipeline.html}} and \textsc{xselect v2.4g} software packages.

To mitigate pile-up effects, spectra were extracted from an annular region defined by two concentric circles with radii of 5 and 15 arcmin. Using the LAXPC gti files corresponding to the high and low flux intervals of $\kappa$ and $\omega$ classes, we extracted SXT spectra in both states. For spectral modelling, orbit-specific ancillary response files (ARFs) were generated using the \textsc {sxt$\_$ARFModule$\_$v02} tool.

\subsection{LAXPC}
LAXPC consists of three independent but identical detectors giving a collecting area of $\sim 6000 cm^2$ at 15 keV, which are proportional counters which can detect X-rays in 3-80 keV energy band. We downloaded the level 1 data of the LAXPC observations from Astrosat Archive \footnote{\url{https://astrobrowse.issdc.gov.in/astro_archive/archive/Home.jsp}} and reduced the data using relevant tools in  LAXPC software v22Aug2015. The pipeline also includes the tools to filter the South Atlantic Anomaly passage and Earth occultation intervals and extract the background subtracted light curves and spectra. For the analysis, we considered only LAXPC 20 as LAXPC 30 was turned off early in the mission and LAXPC 10 indicated abnormal gain variation. 

\begin{table*}[!ht]
\setlength{\tabcolsep}{4pt} 
\centering
\caption{Log of observations}
\renewcommand{\arraystretch}{1.0} 
\label{table1}
\begin{tabular}{ccccccccc}
\hline
\hline
Obs no. & Obs ID    & Observation Date & Class & Effective exposure & Total Peak     & Avg. peak    & Total Dip    & Avg. dip   \\
        & Orbit no. &   (DD-MM-YYYY)   &       &    time (s)        &     time (s)   &  count rate  &     time (s) &  count rate \\
\hline

$\kappa_1$ & 09899 & 27-07-2017 & $\kappa$ & 15535 & 5059   & 11246  & 5385 & 2801 \\
\hline
$\kappa_2$ & 9900/9903  & 27-07-2017 & $\kappa$ & 3123 & 1050  & 11241 & 932 & 2952 \\
  
\hline
$\omega_1$ & 10394 & 30-08-2017 & $\omega$ & 2932 & 1767 & 10727 & 532 & 2771 \\
 
\hline
$\omega_2$ & 10420 & 31-08-2017 & $\omega$ & 10912 & 6903 & 10858 & 1336 & 2799 \\

\hline
\end{tabular}
\begin{flushleft}
Note: Exposures given in total peak time and total dip time columns are used to extract and analyse peak and dip energy spectra, respectively. 
\end{flushleft}
\label{tab1}
\end{table*}
\begin{figure*}[!ht]
    \centering
    \includegraphics[width=0.48\textwidth]{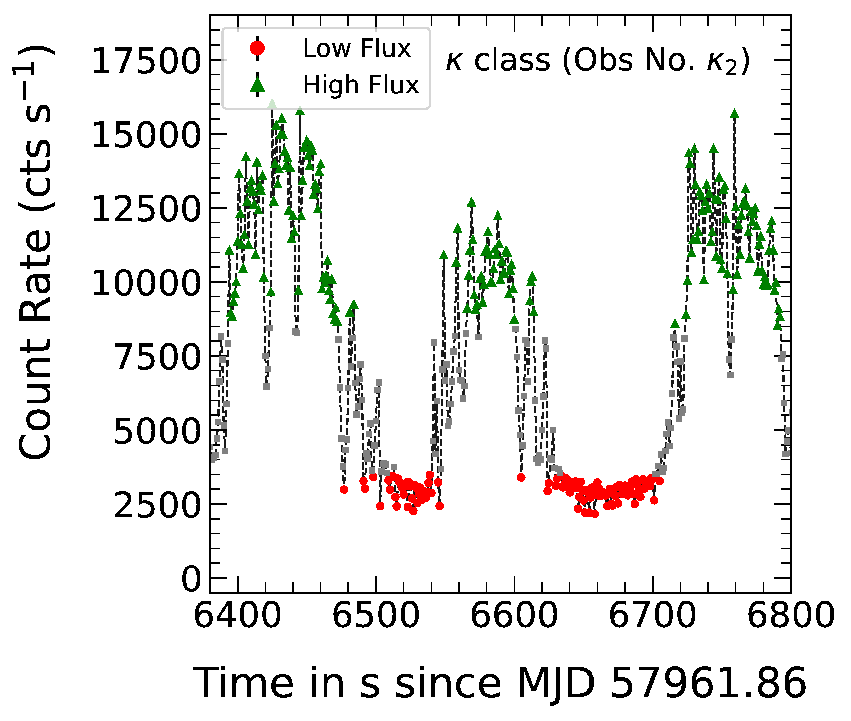}
    \hfill
    \includegraphics[width=0.46\textwidth]{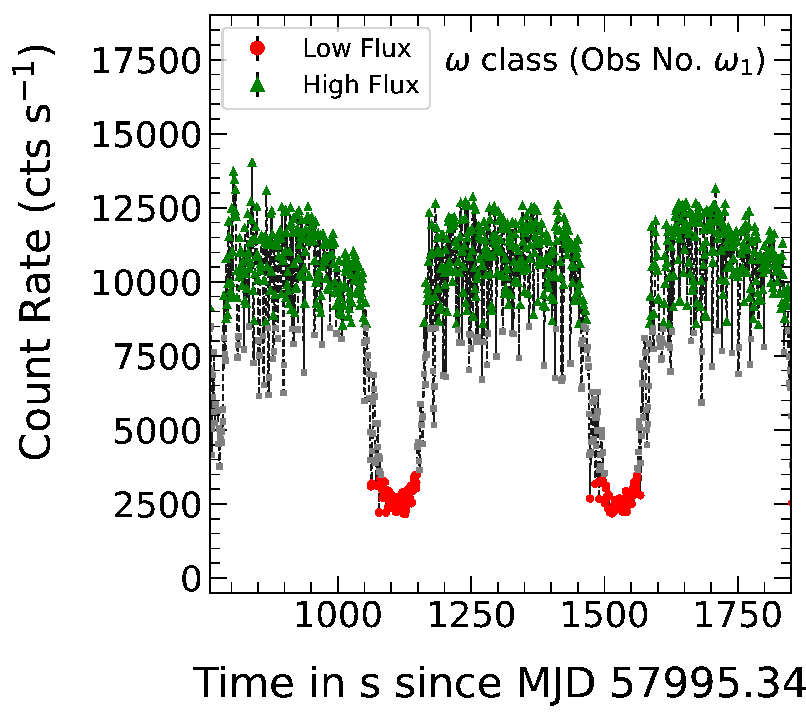}
    \hfill
    \includegraphics[width=0.48\textwidth]{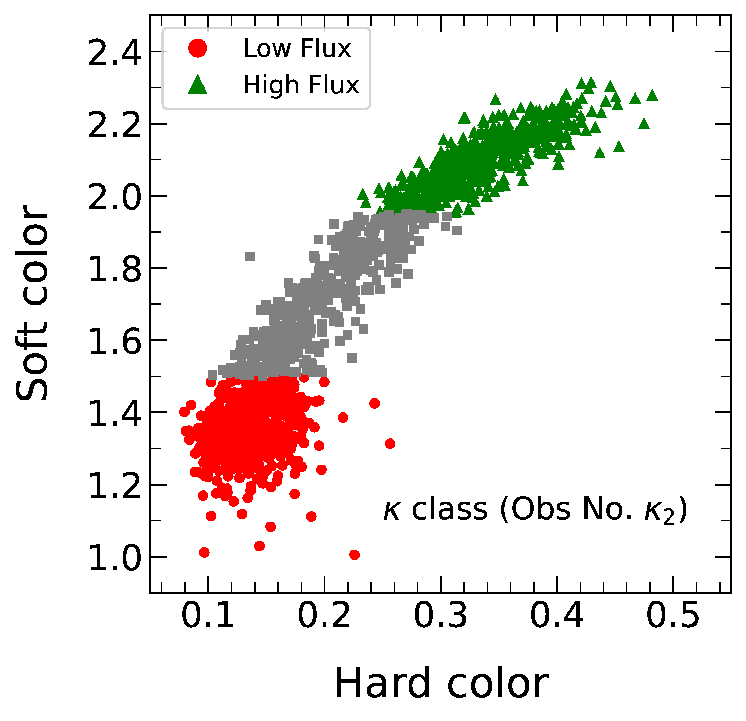}
    \hfill
    \includegraphics[width=0.48\textwidth]{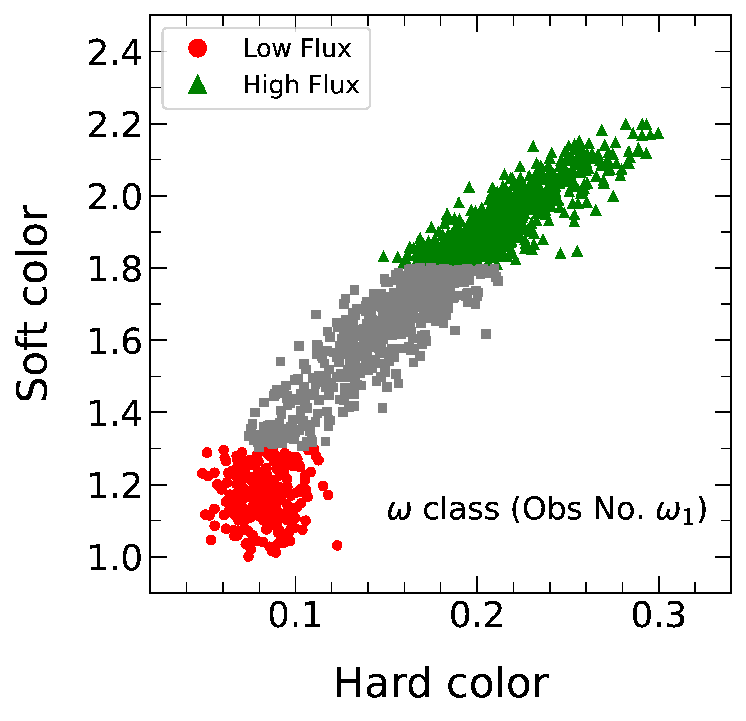}
    \hfill
    \includegraphics[width=0.48\textwidth]{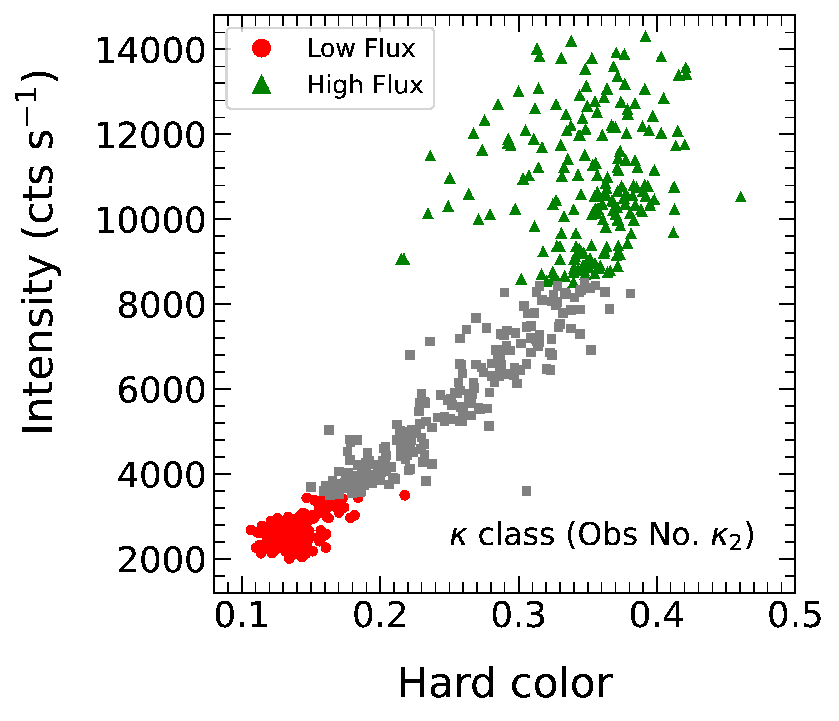}
    \hfill
    \includegraphics[width=0.48\textwidth]{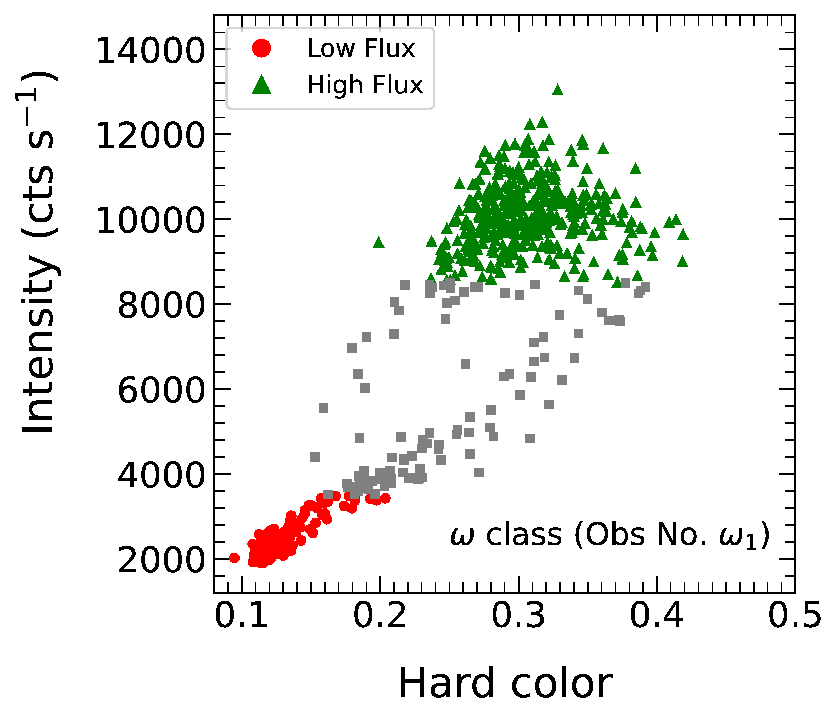}
    \caption{Background-subtracted 3--30.0 keV \asat{}/LAXPC (all units combined) X-ray lightcurve of $\kappa$ class with 1.0 sec time resolution is shown in the top left panel, while the same for the $\omega$ class is shown in the top right panel. The color-color diagrams for the $\kappa$ and $\omega$ class observations are shown in the middle left and middle right panels respectively while the hardness intensity diagrams for the same are shown in the bottom left and bottom right panels, respectively. Selections of high and low flux intervals are shown in green triangles and red squares, respectively in all panels.}
    \label{fig1}
\end{figure*}
\section{Timing analysis and results}\label{sec3}

The background-subtracted LAXPC light curves of GRS~1915+105, shown in top panels of Figure \ref{fig1} reveal the characteristic large-amplitude, quasi-periodic variability that defines the  $\kappa$ (top left panel) and $\omega$ (top right panel) classes. In both $\kappa$ observations, the light curves exhibit slow, asymmetric and irregular transitions between bright, variable, high-flux peaks and long low flux dips, with peak count rates varying between 10000-15000 cts s$^{-1}$ and dip count rates falling below 2500 cts s$^{-1}$. The recurrence time of these oscillations is variable and oscillates in 70-100~s, As observed from Table \ref{tab1}, nearly equal times are spend during high and low flux intervals in $\kappa$ class.
On the other hand, the $\omega$ class observations display a regular, cyclic and smooth intensity variations but with longer burst durations and broader high-flux plateaus, producing a smoother oscillatory morphology in which the high-flux intervals are substantially extended relative to the dips. The source spends time nearly three times longer during high flux compared to that of the low flux intervals as observed from Table \ref{tab1}. The average count rates during the bright phases of the $\omega$ class remain stable around 11000~cts s$^{-1}$, while the dips reach values comparable to those observed in the $\kappa$ class. The high and low flux intervals are identified in Table \ref{tab1} and correspond to the most stable segments of these oscillations and identified with green triangles and red circles respectively.

For better distinction of high and low flux regimes, we calculated color-color diagrams (CCD) and hardness intensity diagrams (HID) of both $\kappa$ and $\omega$ class observations and show them in the middle (left and right) and bottom (left and right) panels of Figure \ref{fig1}, respectively. Following \citet{Dhaka2025}, the hard and soft color are defined as the ratio of the 10-30 keV to 3-6 keV count rate while the soft color is defined as the 6-10 keV to 3-6 keV count rate. The intensity is defined as the 3-30 keV count rate. Both CCDs and HIDs show distinct high and low flux intervals. For better visibility, our selection of high and low flux intervals is marked by green triangles and red circles in all panels of Figure \ref{fig1} respectively. Average count rate, total high and low-flux time intervals during each observation, which are used for further flux-resolved spectral analysis during high and low fluxes, are provided in Table \ref{tab1}. 
The clear separation between the peak and dip intensities in both classes demonstrates that the source undergoes rapid and well-defined transitions in accretion state, motivating the detailed spectral and covariance studies presented in the later sections.


\section{Spectral Analysis and Results}\label{sec4}

To investigate the spectral evolution associated with the rapid high-to-low flux transitions in the $\kappa$ and $\omega$ variability classes of GRS~1915+105, we performed joint broadband spectral modelling using \textit{AstroSat}/SXT (0.8--7 keV) and LAXPC20 (3--20 keV) data. For each observation, flux-resolved spectra were extracted from the high-flux (burst) and low-flux (dip) intervals identified using suitable GTI files in the background-subtracted light curves (see Figure \ref{fig1}). 

An important step is the choice of spectral models to describe burst and dip spectra. We start with single-component models like diskbb \citep{Mitsuda1984,Makishima1986}, powerlaw, and nthcomp \citep{Zdziarski1996}. While fitting burst and dip spectra, we find that single-component models provide unacceptably large reduced $\chi^2$. In order to improve the fit, we use two-component models like diskbb+powerlaw, THComp*Diskbb. We also tried a Gaussian model component at 6.4 keV. We find that among all two-component models, a simple, physical model, which is a combination of a multi-temperature disc blackbody (diskbb) in XSpec version 12.14.1 \citep{Arnaud1996} and a convolved thermal Comptonization continuum model \citep[Thcomp;][]{Zdziarski2020} in XSpec, along with a physical disc reflection model \citep[Xillver;][]{Garcia2014, Dauser2014} in \textsc{XSpec}, can adequately describe both burst and dip spectra for all observations. The THComp model assumes photons from the innermost disk are seed photons for Comptonization \citep{Zdziarski2020} with a covering fraction that determines the fraction of seed photons that will be comptonised. Additionally, an X-ray absorption cross-section, modelled as the Tuebingen-Boulder ISM absorption \citep[TBabs;][]{Wilms2000} was applied to correct the interstellar and local absorption effects along the line-of-sight by utilizing the cross section from \citet{Verner1996}, and abundances from \citep{Wilms2000}.

The spectra were fitted simultaneously using the model \texttt{TBabs*TBabs(thcomp*diskbb + xillver)}, which accounts for interstellar and local absorption, a multicolour disc blackbody, a Comptonized continuum, and a weak reflection component. A constant factor was included to correct for SXT and LAXPC cross-calibration differences.

To determine the degree of physical change between the high- and low-flux intervals, the low-flux spectra were fitted sequentially under four different constraints and shown in Table \ref{tab2}. In Case A, only the continuum normalisations were allowed to vary relative to the high-flux spectrum, while in Case B, both the normalisations and the local absorption column density were free. Table \ref{tab2} shows that low flux spectra cannot be fitted with Case A and Case B. 
Whereas, Case C allowed the Comptonization photon index to vary as well. This hierarchy enabled us to identify the minimum set of parameters required to explain the strong spectral softening and hardening observed during the rapid transitions.

\begin{figure*}[!ht]
    \centering
    \includegraphics[width=0.37\textwidth,angle=270]{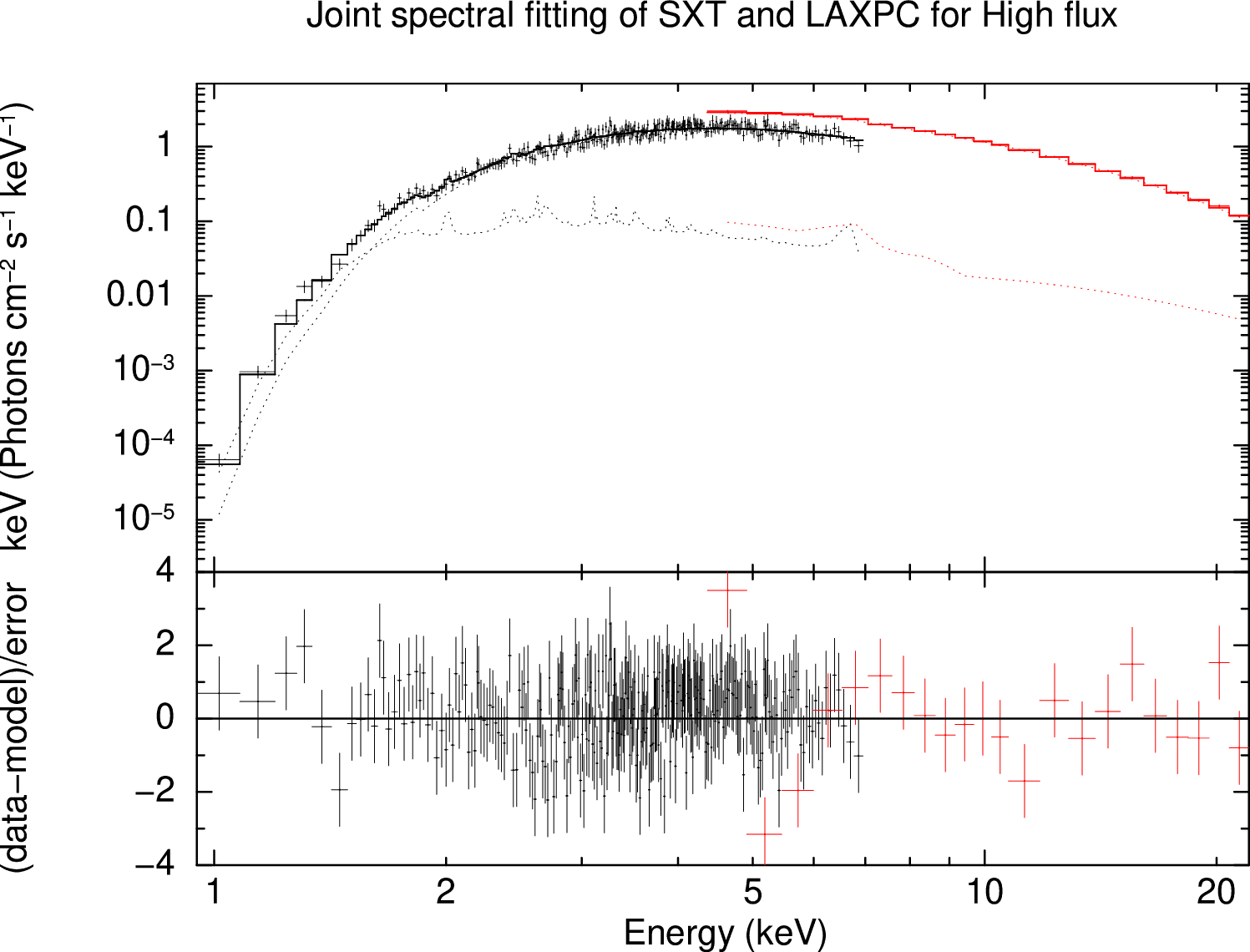}
    \hfill
    \includegraphics[width=0.37\textwidth,angle=270]{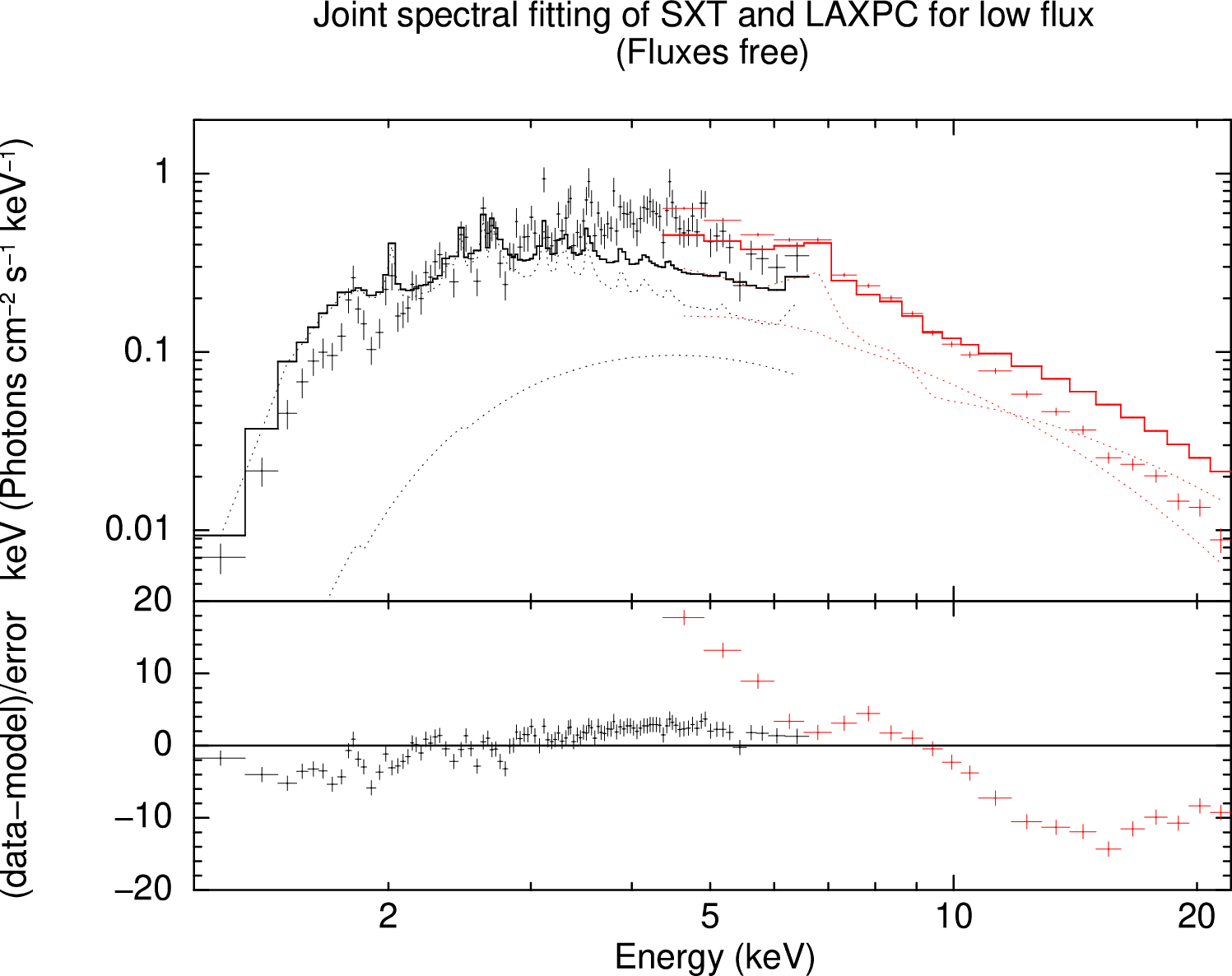}
    \hfill
    \includegraphics[width=0.37\textwidth,angle=270]{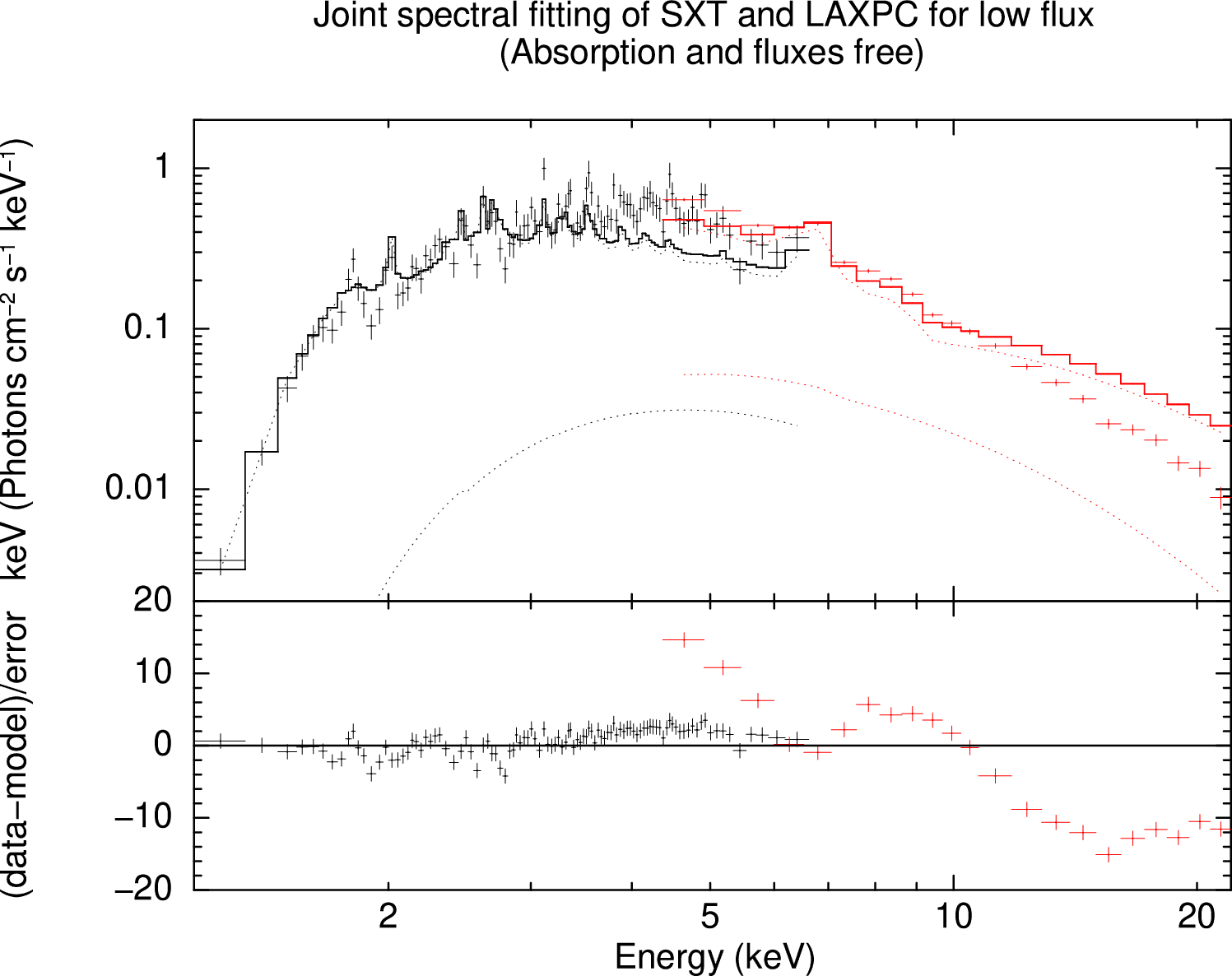}
    \hfill
    \includegraphics[width=0.37\textwidth,angle=270]{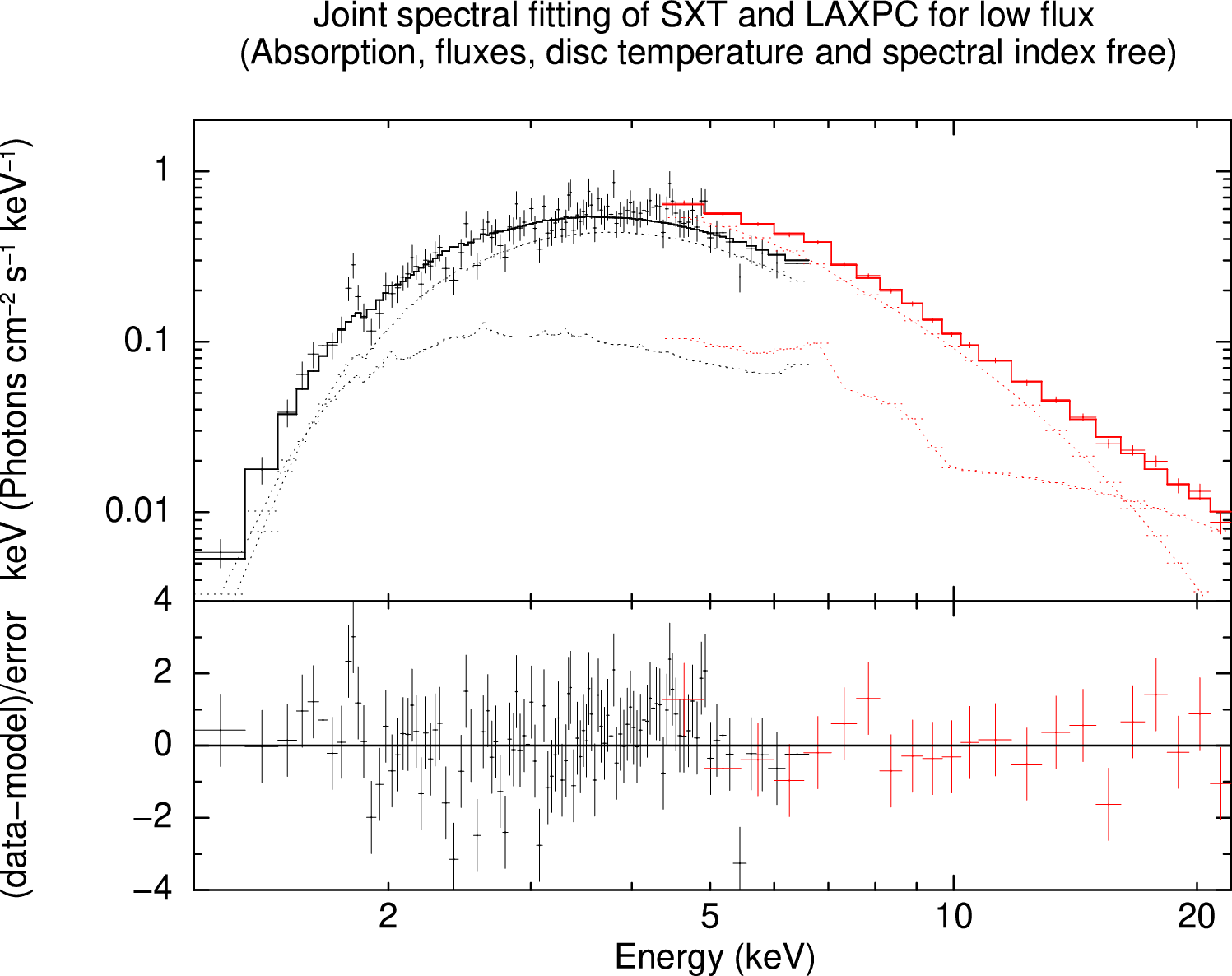}
    
    \caption{Unfolded \asat{} SXT/LAXPC20 mean spectra of \src{} in the 1.0–25.0 keV energy range. The upper-left panel shows the best-fit joint spectra for the high-flux state of the $\kappa$ class (Obs. $\kappa_2$), including the model components and data-to-model ratio. The upper-right, and lower-left panels display the joint spectral fits for Case-A, and Case-B, respectively, corresponding to the low-flux state of the $\kappa$ class (Obs. $\kappa_2$). The lower-right panel presents the best-fit joint spectra for the low-flux (Case-C) of the $\kappa$ class.}
    \label{fig2}
\end{figure*}
\begin{table}[!ht]
\centering
\caption{Spectral modelling results for different cases for $\kappa$ and $\omega$ classes low flux state. }
\begin{tabular}{cccc}
\hline
\hline
 Obs.      & \multicolumn{3}{c}{$\chi^2/Dof$} \\
\cline{2-4}
State      &  Case-A         &  Case-B         & Case-C  \\
\hline
$\kappa_1$ &  2073/129      &   2040/128       & 130/126 \\
$\kappa_2$ &  2354/124      &   2097/123       & 148/121 \\
$\omega_1$ &  1802/191      &   1381/190       & 186/188 \\
$\omega_2$ &  1178/136      &   828/135        & 122/133 \\
\hline
\end{tabular}
\begin{flushleft}
\textbf{Note.} Case A: When only normalisations are kept free, Case B: When normalisations and local absorption are kept free,  Case C: When normalisations, absorption, disc temperature, and power-law spectral index are kept free to vary during fitting. 
\end{flushleft}
\label{tab2}
\end{table}
\subsection{Broadband Spectra of the $\kappa$ Class}

\subsubsection{High-flux interval}

The unfolded broadband spectra for the $\kappa$ class, shown in Fig \ref{fig2} for Obs.~$\kappa_1$ and $\kappa_2$,  indicate that the source is in a bright, disc-dominated state during the high-flux interval. The accretion disc displays a high colour temperature of approximately 3.8--4.2\,keV, while the Comptonized tail is steep with $\Gamma_\tau$ between about 3.1 and 3.4. The reflection component remains moderately ionised, with  $\log\xi$ close to 4, and the 0.3--30\,keV absorbed flux lies near $(2.3$--$2.5)\times10^{-8}\,{\rm erg\,s^{-1}\,cm^{-2}}$. These characteristics are consistent with the disc providing a substantial fraction of the seed photons that efficiently cool the corona during the high-intensity phases.

\subsubsection{Low-flux interval}

During the low-flux intervals of the $\kappa$ class, the spectral shape changes markedly, as illustrated in the upper-right and middle-right panels of Fig \ref{fig2}. When only the normalisations were allowed to vary (Case A), the fits were statistically unacceptable, with $\chi^{2}/{\rm dof}$ values approaching 2000. This immediately shows that the low-flux spectrum cannot be produced by a simple rescaling of the high-flux spectrum components. Even allowing the absorption to vary (Cases B) does not greatly improve the quality of the fits, and a satisfactory fit is ultimately achieved only when the Comptonization photon index is also allowed to change (Case C). 

The best-fitting models reveal that the disc temperature decreases sharply from values near 4 keV during the high-flux interval to approximately 2.3--2.7 keV during the low-flux interval. At the same time, the Comptonized continuum becomes noticeably harder, with the photon index decreasing to about 2.6--2.7. The total absorbed flux drops correspondingly by a factor of four to five relative to the high state. These changes strongly suggest that the low-flux intervals in the $\kappa$ class are produced by a rapid cooling or partial recession of the inner accretion disc, together with a hardening of the corona caused by the reduced availability of soft seed photons.

\begin{table*}[!ht]
\setlength{\tabcolsep}{5pt}
\centering
\caption{Best-fit spectral parameters for the broadband X-ray spectral modelling for $\kappa$ spectral class of \src{} using \texttt{TBabs*TBabs(thcomp*diskbb + xillver)}.}
\renewcommand{\arraystretch}{1.5}
\begin{tabular}{cccccccc}
\hline\hline
Component & Parameter & Unit & High flux      & Low flux ($\kappa_1$) & High flux      & Low flux ($\kappa_2$)\\
          &           &      & ($\kappa_1$)   & CASE-C                & ($\kappa_2$)   & CASE-C\\
\hline
\texttt{TBabs}      &     $N_{\rm H}$      &    $10^{22}$ cm$^{-2}$    & $4.14_{-0.54}^{+0.55}$    & $3.00_{-0.27}^{+0.32}$  & $3.69_{-0.46}^{+0.41}$   & $2.68_{-0.24}^{+0.27}$\\
\texttt{thcomp}     & $\Gamma_{\tau}$      &                           & $3.18_{-0.51}^{+0.64}$    & $2.65_{-0.14}^{+0.15}$  & $3.40_{-0.45}^{+0.60}$   & $2.66_{-0.13}^{+0.12}$\\
                    &    $kT_e$            &              keV          & $0.66_{-0.10}^{+0.12}$    &       $0.66^\star$      & $0.67_{-0.05}^{+0.14}$   &       $0.67^\star$     \\
                    & $C_{\rm f}$          &                           & $0.80_{-0.13}^{+0.07}$    &       $0.80^\star$      & $0.85_{-0.13}^{+0.10}$   &       $0.85^\star$        \\
\texttt{diskbb}     & $T_{\rm in}$         &              keV          & $3.87_{-0.23}^{+0.57}$    & $2.77_{-0.05}^{+0.06}$  & $4.17_{-0.43}^{+0.64}$   & $2.60_{-0.06}^{+0.07}$ \\
 \texttt{xillver}   & $\log\xi$            & erg cm s$^{-1}$           & $4.01_{-0.55}^{+0.42}$    &        $4.01^\star$     & $4.09_{-0.73}^{+0.38}$   &        $4.09^\star$       \\
                    & Norm                 &                           & $0.50_{-0.25}^{+1.04}$    & $0.04_{-0.02}^{+0.04}$  & $0.67_{-0.45}^{+0.52}$   & $0.04_{-0.02}^{+0.03}$\\
\texttt{Constant}   &   $ \rm C_{LAXPC}$   &                           & $1.08_{-0.03}^{+0.03}$    &  $1.22_{-0.07}^{+0.08}$ & $1.53_{-0.45}^{+0.52}$   &  $1.30_{-0.06}^{+0.07}$\\
\hline
\texttt{Flux$_{0.3-30}$} &          & $10^{-9}$ erg s$^{-1}$ cm$^{-2}$ & $24.7_{-0.1}^{+0.1}$      & $5.81_{-0.05}^{+0.05}$ & $23.3_{-0.1}^{+0.1}$      & $4.81_{-0.05}^{+0.05}$    \\
\texttt{$\chi^2$/dof} &                    &                           & 297/283                   & 130/126                & 281/282                   & 148/121           \\
\hline
\end{tabular}
\begin{flushleft}
\textbf{Note.} $N_{\rm H}$ is the local Hydrogen column density; $\Gamma_{\tau}$ and $kT_e$ are the thcomp photon index and electron temperature, respectively; $C_{\rm f}$ is the scattering fraction; $T_{\rm in}$ is the disk inner temperature; $\xi$ is the ionisation parameter of the reflector. Uncertainties correspond to 90\% confidence ($\Delta\chi^2=2.706$). \\
$\star$: The parameter is kept fixed. \\
Flux$_{0.3-30}$: Absorbed X-ray flux in 0.3-30.0 keV energy range.
\end{flushleft}
\label{tab3}
\end{table*}


\subsection{Broadband Spectra of the $\omega$ Class}

\subsubsection{High-flux interval}
\begin{figure*}[!ht]
    \centering
    \includegraphics[width=0.37\textwidth,angle=270]{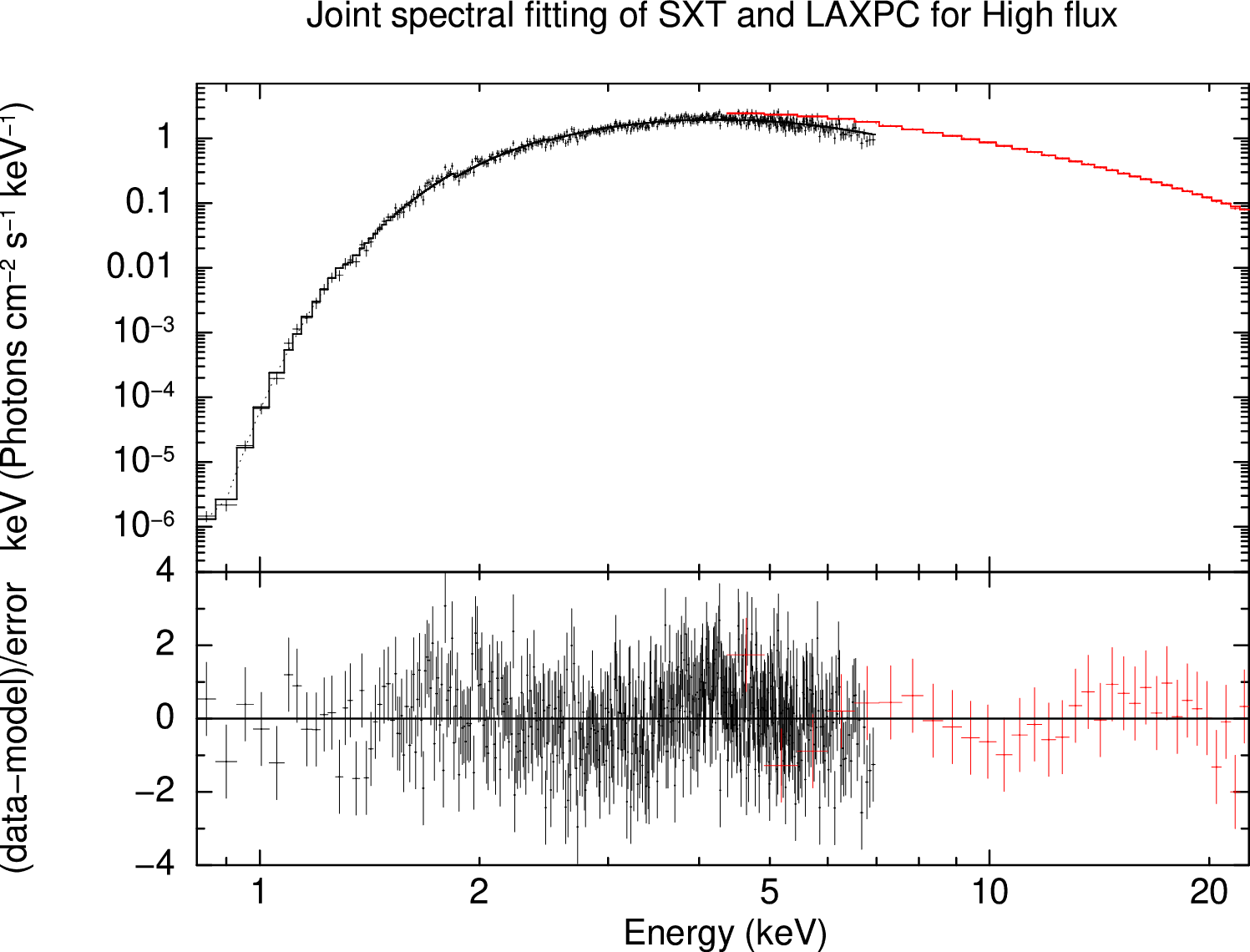}
    \hfill
    \includegraphics[width=0.37\textwidth,angle=270]{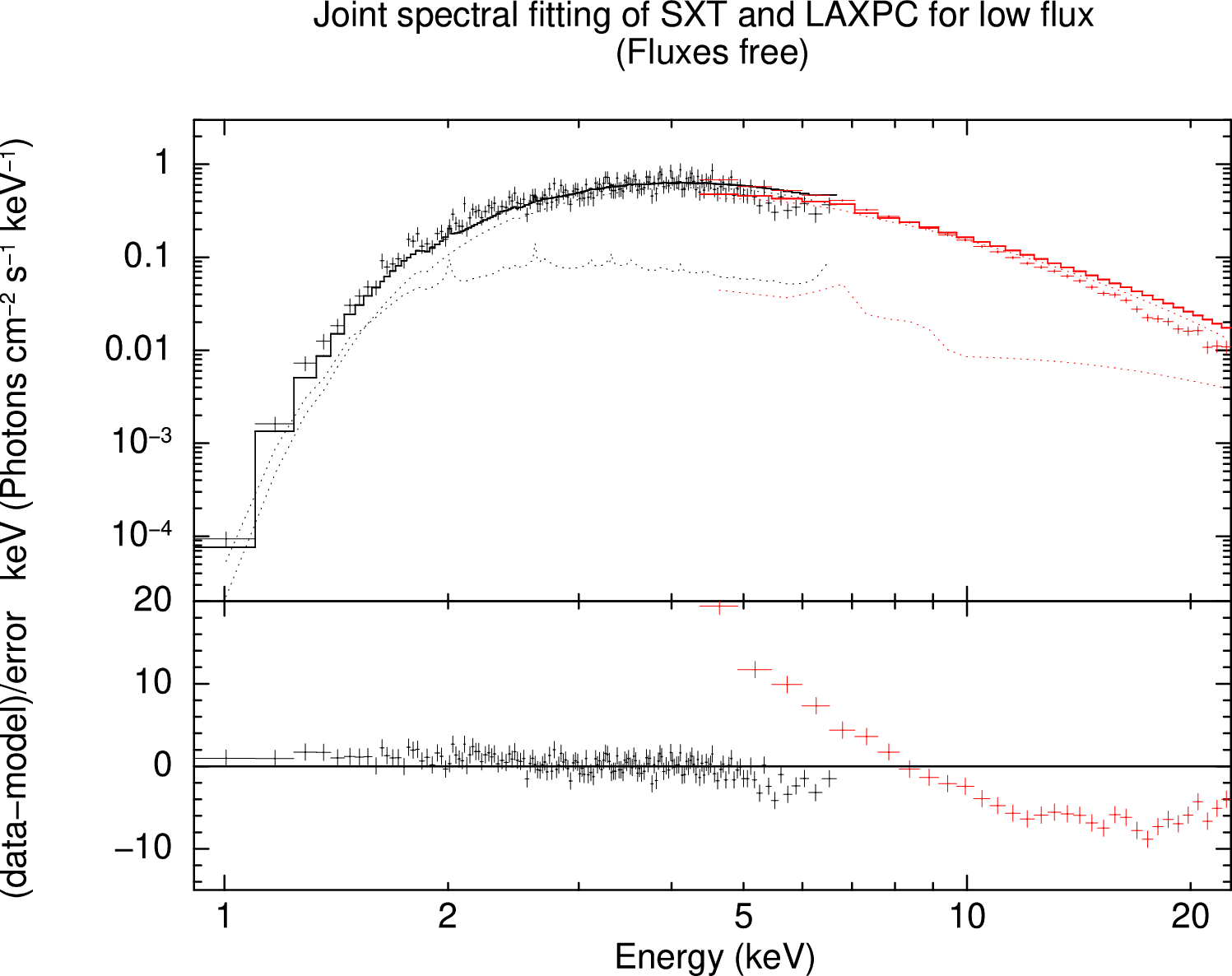}
    \hfill
    \includegraphics[width=0.37\textwidth,angle=270]{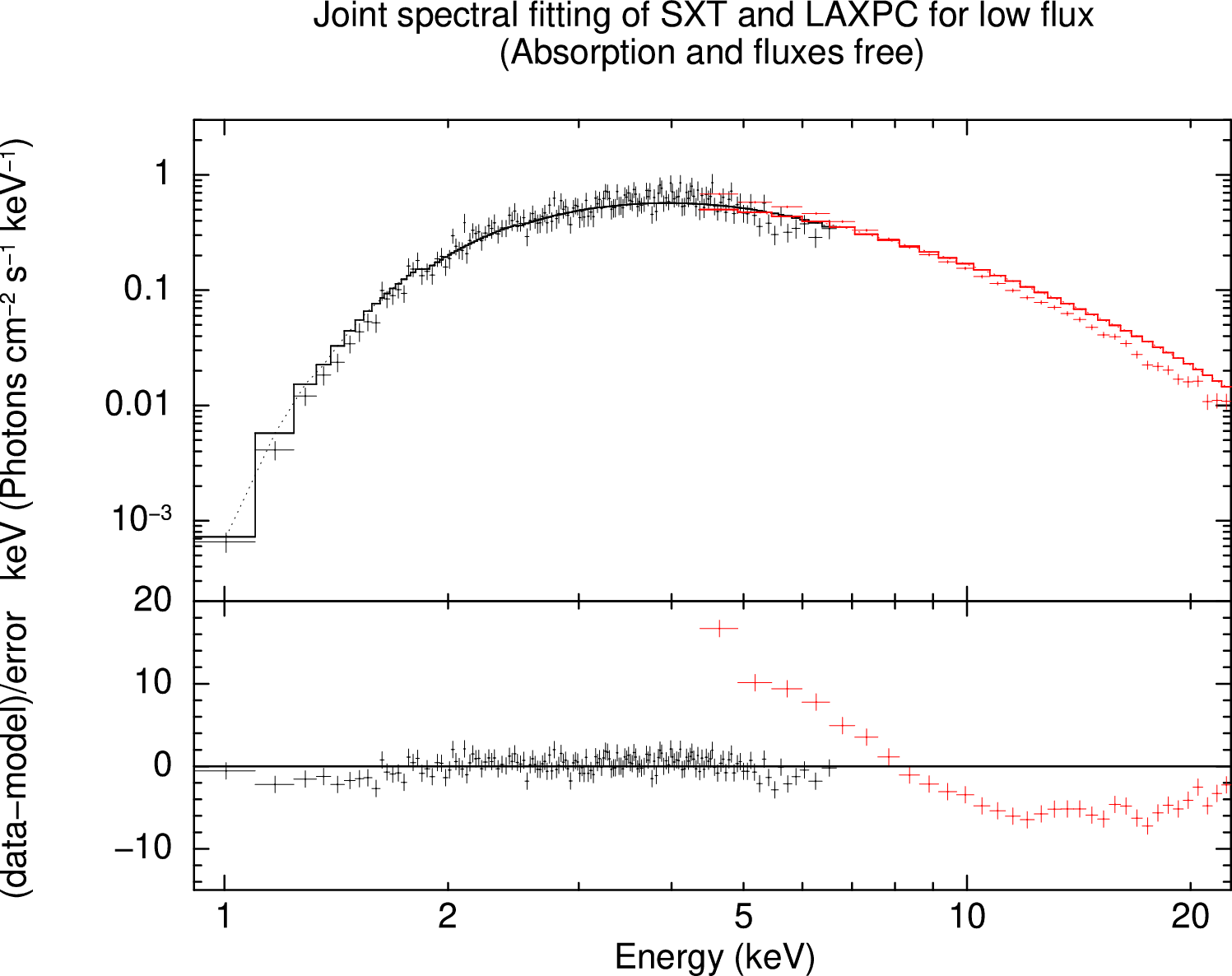}
    \hfill
    \includegraphics[width=0.37\textwidth,angle=270]{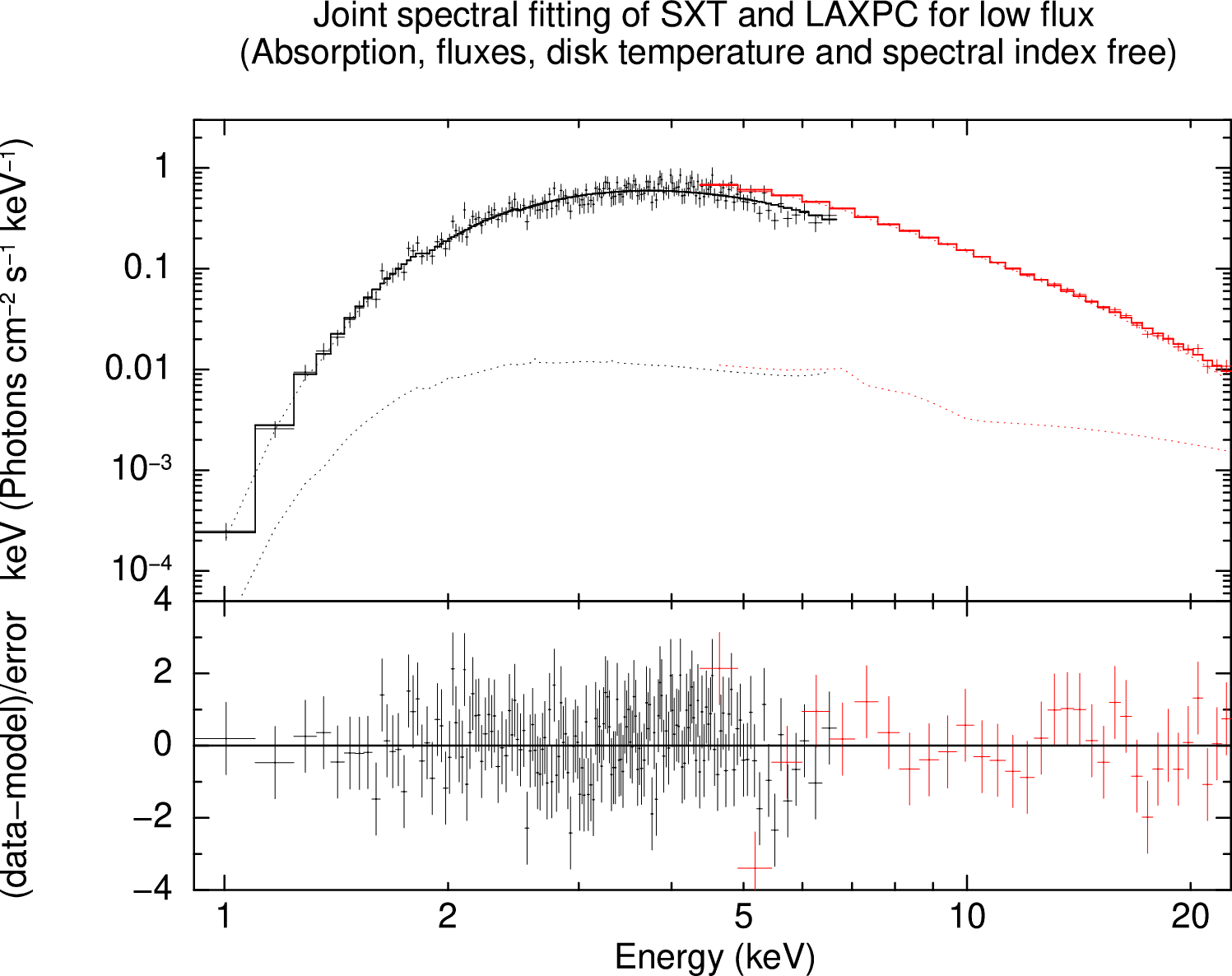}
    \caption{Unfolded \asat{} SXT/LAXPC20 mean spectra of \src{} in the 1.0–25.0 keV energy range. The upper-left panel shows the best-fit joint spectra for the high-flux state of the $\omega$ class (Obs. $\omega_1$), including the model components and data-to-model ratio. The upper-right and middle-left panels display the joint spectral fits for Case-A and Case-B, respectively, corresponding to the low-flux state of the $\omega$ class (Obs. $\omega_1$). The lower-right panel presents the best-fit joint spectra for the low-flux (Case-C) of the $\omega$ class.}
    \label{fig3}
\end{figure*}

The high-flux spectra of the $\omega$ class, displayed in Figure~\ref {fig3} for Obs.~$\omega1$ and $\omega2$, are broadly similar to those of the $\kappa$ class, although the disc temperatures in this case tend to lie between 4.2 and 4.6 keV. 
\begin{figure*}[!ht]
    \centering
    \includegraphics[width=0.49\textwidth]{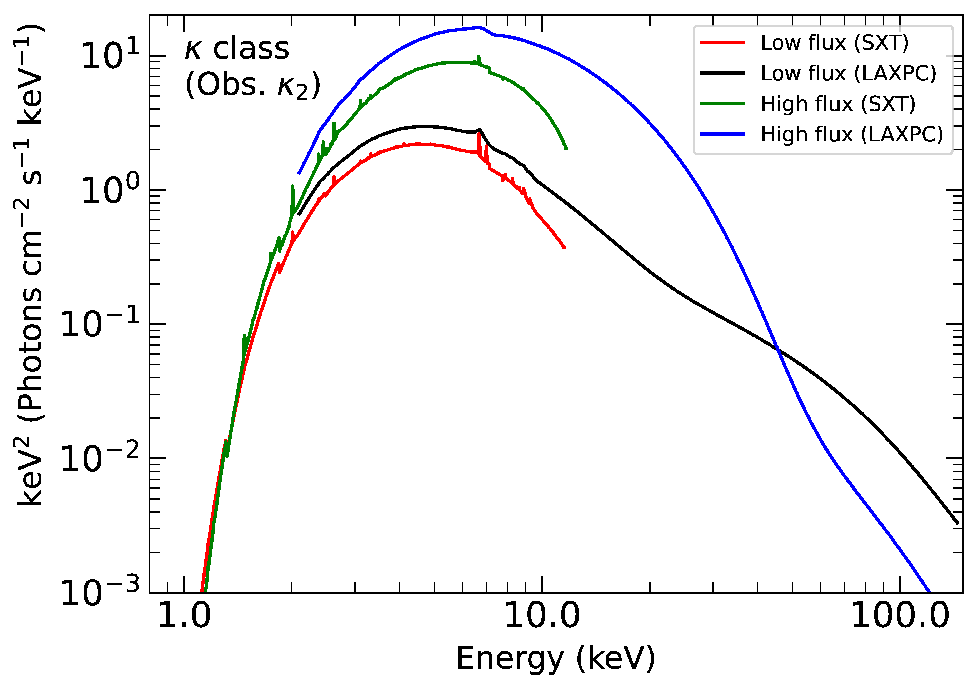}
    \hfill
    \includegraphics[width=0.49\textwidth]{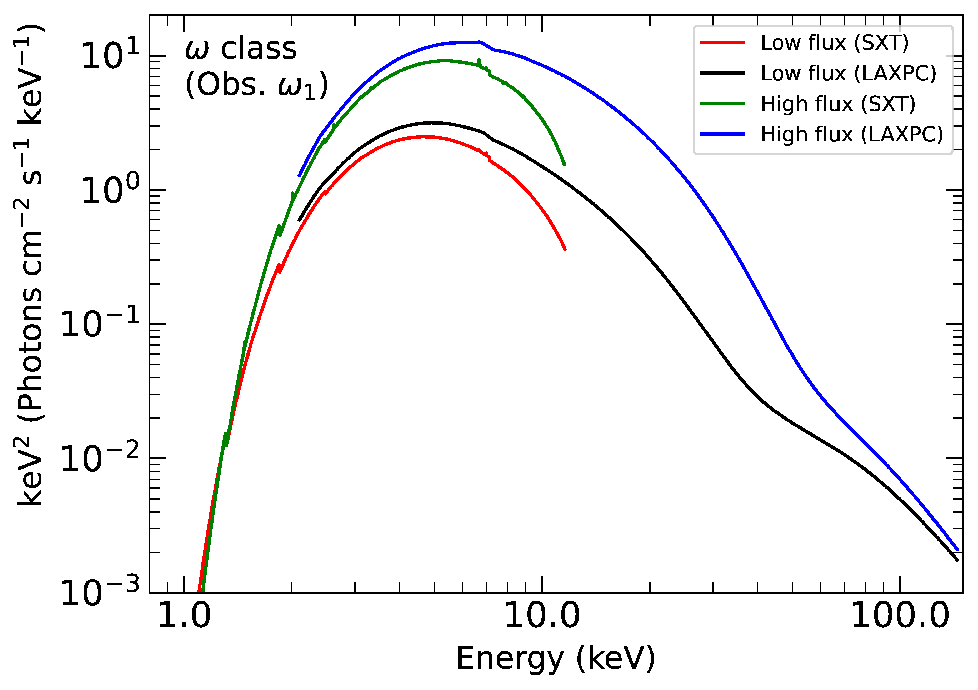}
    \caption{The left panel presents the best-fit continuum models for both the high and low-flux observations of the $\kappa$ class (Obs. $\kappa_2$),  while the right panel shows the best-fit continuum models for both the high and low-flux observations of the $\omega$ class (Obs. $\omega_1$). }
    \label{fig3a}
\end{figure*}
The Comptonized continuum is steep with photon indices between roughly 2.6 and 3.4, and the reflection component remains highly ionised, with $\log\xi$ generally in the range 3.7--4.5. The absorbed flux, spanning $(1.8$--$2.5)\times10^{-8}\,{\rm erg\,s^{-1}\,cm^{-2}}$, again indicates that the source is in a bright, disc-dominated configuration during the high-flux segments of the $\omega$ class.

\subsubsection{Low-flux interval}

When the source transitions into its low-flux state in the $\omega$ class, the spectral shape again changes significantly. As in the $\kappa$ class, the low-flux spectra cannot be described by uniform rescaling, and acceptable fits are obtained only when the disc temperature and Comptonization photon index are allowed to vary. The best-fitting models show that the disc temperature declines from its high-flux values of approximately 4.2--4.6 keV to about 2.8--3.3\,keV in Obs. $\omega1$ and to roughly 2.84--2.86 keV in Obs.~$\omega2$. The Comptonized continuum also hardens, with photon indices falling to values between about 2.1 and 2.7. The absorbed flux simultaneously declines to approximately $5\times10^{-9}\,{\rm erg\,s^{-1}\,cm^{-2}}$, indicating a substantial reduction in luminosity during the dip. These trends confirm that the cooling and recession of the disc, along with diminished coronal cooling, play a major role in shaping the low-flux intervals in the $\omega$ class as well.

Best-fit models for both $\kappa$ and $\omega$ class spectral analysis are shown in the left and right panels of Figure \ref{fig3a}.
\begin{table*}[!ht]
\setlength{\tabcolsep}{4pt}
\centering
\caption{Best-fit spectral parameters for the broadband X-ray spectral modelling for $\omega$ spectral class of \src{} using \texttt{TBabs*TBabs(thcomp*diskbb + xillver)}.}
\renewcommand{\arraystretch}{1.4}
\begin{tabular}{cccccccc}
\hline\hline
Component & Parameter & Unit & High flux      & Low flux ($\omega_1$) & High flux     & Low flux ($\omega_2$)\\
          &           &      & ($\omega_1$)   & CASE-C                & ($\omega_2$)  & CASE-C\\

\hline
\texttt{TBabs}      &     $N_{\rm H}$      &    $10^{22}$ cm$^{-2}$    & $2.93_{-0.17}^{+0.24}$    & $1.79_{-0.13}^{+0.14}$   & $3.76_{-0.24}^{+0.22}$        & $2.63_{-0.24}^{+0.26}$\\
\texttt{thcomp}     & $\Gamma_{\tau}$      &                           & $2.65_{-0.15}^{+0.17}$    & $2.13_{-0.04}^{+0.08}$   & $3.43_{-0.11}^{+0.49}$        & $2.69_{-0.13}^{+0.11}$  \\
                    &    $kT_e$            &              keV          & $0.76_{-0.05}^{+0.04}$    &       $0.76^\star$       &   $0.66_{-0.03}^{+0.03}$      &       $0.66^\star$        \\
                    & $C_{\rm f}$          &                           & $0.90^{\star}$            &       $0.90^\star$       &       $0.84^\star$            &       $0.84^\star$        \\
\texttt{diskbb}     & $T_{\rm in}$         &              keV          & $4.60_{-0.08}^{+0.09}$    & $3.77_{-0.14}^{+0.11}$   & $4.21_{-0.04}^{+0.03}$        & $2.86_{-0.06}^{+0.04}$  \\
 \texttt{xillver}   & $\log\xi$            & erg cm s$^{-1}$           & $3.72_{-0.49}^{+0.66}$    &        $3.72^\star$      &     $>4.52$                   &        $4.52^\star$       \\
                    & Norm                 &                           & $0.03_{-0.02}^{+0.05}$    & $0.001_{-0.001}^{+0.002}$& $0.42_{-0.14}^{+0.15}$        & $0.03_{-0.01}^{+0.02}$ \\
\texttt{Constant}   &   $ \rm C_{LAXPC}$   &                           & $1.13_{-0.01}^{+0.01}$    &  $1.23_{-0.04}^{+0.04}$  &  $1.46_{-0.02}^{+0.02}$       &  $1.46_{-0.06}^{+0.06}$  \\

\hline
\texttt{Flux$_{0.3-30}$} &          & $10^{-9}$ erg s$^{-1}$ cm$^{-2}$ & $24.7_{-0.5}^{+0.5}$      & $5.93_{-0.04}^{+0.04}$   & $18.6_{-0.06}^{+0.06}$        & $4.93_{-0.04}^{+0.04}$   \\
\texttt{$\chi^2$/dof} &                    &                           & 531/497                   & 186/188                  & 404/328                       & 122/133           \\
\hline
\end{tabular}
\begin{flushleft}
\textbf{Note.} $N_{\rm H}$ is the local Hydrogen column density; $\Gamma_{\tau}$ and $kT_e$ are the thcomp photon index and electron temperature, respectively; $C_{\rm f}$ is the scattering fraction; $T_{\rm in}$ is the disk inner temperature; $\xi$ is the ionisation parameter of the reflector. Uncertainties correspond to 90\% confidence ($\Delta\chi^2=2.706$). \\
$\star$: The parameter is kept fixed. \\
Flux$_{0.3-30}$: Absorbed X-ray flux in 0.3-30.0 keV energy range.
\end{flushleft}
\label{tab4}
\end{table*}

\subsection{Parameter Evolution Across the Two Classes and spectral behaviour}
A comparison of the results summarised in Tables \ref{tab2}, \ref{tab3} and \ref{tab4} shows that both the $\kappa$ and $\omega$ classes exhibit the same qualitative spectral evolution during their rapid high--to--low transitions. In every observation, the inner disc temperature decreases by roughly 1--2\,keV when the source enters the low-flux interval.

This cooling is accompanied by a noticeable hardening of the Comptonized emission. Table \ref{tab4} shows that during the transition from high to low fluxes in the $\omega$ class, photon powerlaw index decreases from 2.65$^{+0.17}_{-0.15}$ to 2.13$^{+0.08}_{-0.04}$ during $\omega_1$ observation, while during $\omega_2$ observation the powerlaw index decreases from 3.43$^{+0.49}_{-0.11}$ to 2.69$^{+0.11}_{-0.13}$. Similarly, Table \ref{tab3} shows a change in photon powerlaw index in $\kappa_2$ class from 3.40$^{+0.60}_{-0.45}$ during high flux to 2.66$^{+0.12}_{-0.13}$ during low flux, although the change is not significant during the transition of $\kappa_1$ observation. We do not observe significance change in the electron temperature or covering fraction possibly due to the limited spectral coverage and resolution.  
The ionisation of the reflector, however, remains relatively stable across the transition, implying that the most significant structural changes occur within the disc and corona rather than in the reflecting medium.
The fact that Case C consistently provides the best statistical fit confirms that variations in the disc temperature in all cases and the spectral index of the Comptonized continuum in three out of four cases are essential components of the transition and cannot be neglected.

The combined SXT+LAXPC spectral analysis therefore shows that the high-flux intervals of both the  $\kappa$ and $\omega$ classes correspond to bright, disc-dominated states in which a hot inner disc supplies an abundance of seed photons that cool the corona efficiently. During the low-flux intervals, the inner disc cools and possibly recedes, and the reduced soft-photon flux leads to a harder Comptonized spectrum. These changes result in a reduction of the observed flux by factors of four to five over timescales of only a few tens of seconds. The spectral evolution revealed by our analysis strongly supports the view that the 
$\kappa$ and $\omega$ classes arise from rapid radiation-pressure-driven limit-cycle oscillations in which the inner disc undergoes repeated cycles of evacuation and refilling, accompanied by corresponding changes in the coronal cooling efficiency.

\section{Covariance Analysis and Results}\label{sec5}
In order to investigate the origin of the rapid spectral variability in the $\kappa$ and $\omega$ classes, we complemented the time-averaged spectral study with a covariance spectral analysis. The covariance spectrum measures the amplitude of variability in each energy channel that is linearly correlated with variations in a broad reference band. It therefore provides a powerful means to isolate the spectral components that drive the fast flux changes, even when these components contribute only a fraction of the time-averaged emission. Following the prescription of \citep{Wilkinson2009}, we computed covariance spectra and error on covariance for both the high- and low-flux intervals of each observation by using 
\begin{equation}
    \label{eqn:cov}
    \centering
    \sigma^2_{cov} = \dfrac{1}{N-1}\sum^N_{i=1}(X_i-\overline{X})(Y_i-\overline{Y})
\end{equation}
Where $Y_i$ is the count rate of the ${i}$th time bin in a reference band (3-20 keV here), the reference band is chosen such that the signal-to-noise ratio (SNR) is very high. 
To calculate covariance, we choose energy ranges: 3-4, 4-5, 5-6, 6-7, 7-8, 8-9, 9-10,10-11,11-13, 13-15, 15-20 keV to make sure that SNR is sufficiently high so that covariance errors are constrained. For each energy range, lightcurves are extracted with a time bin size of 0.1 sec so that the Nyquist frequency is 5 Hz. Typical segment lengths are of the order of hundreds of seconds (see Table \ref{tab1}). Therefore, the covariance spectra computed and used in our analysis are confined within the Fourier frequency range of 0.01-5 Hz. 

$\sigma^2_{cov}$ is again averaged over all the segments for each energy bin and plotted against energy to generate the covariance spectra. In regions where the signal-to-noise ratio of the other light curves is low, the count rates will be averaged to give a null contribution. Additionally, the spectra highlight regions where the signal variability from the specific energy bin is correlated with the signal variability from the reference band.

\begin{figure*}[!ht]
    \centering
    \includegraphics[width=0.35\textwidth,angle=270]{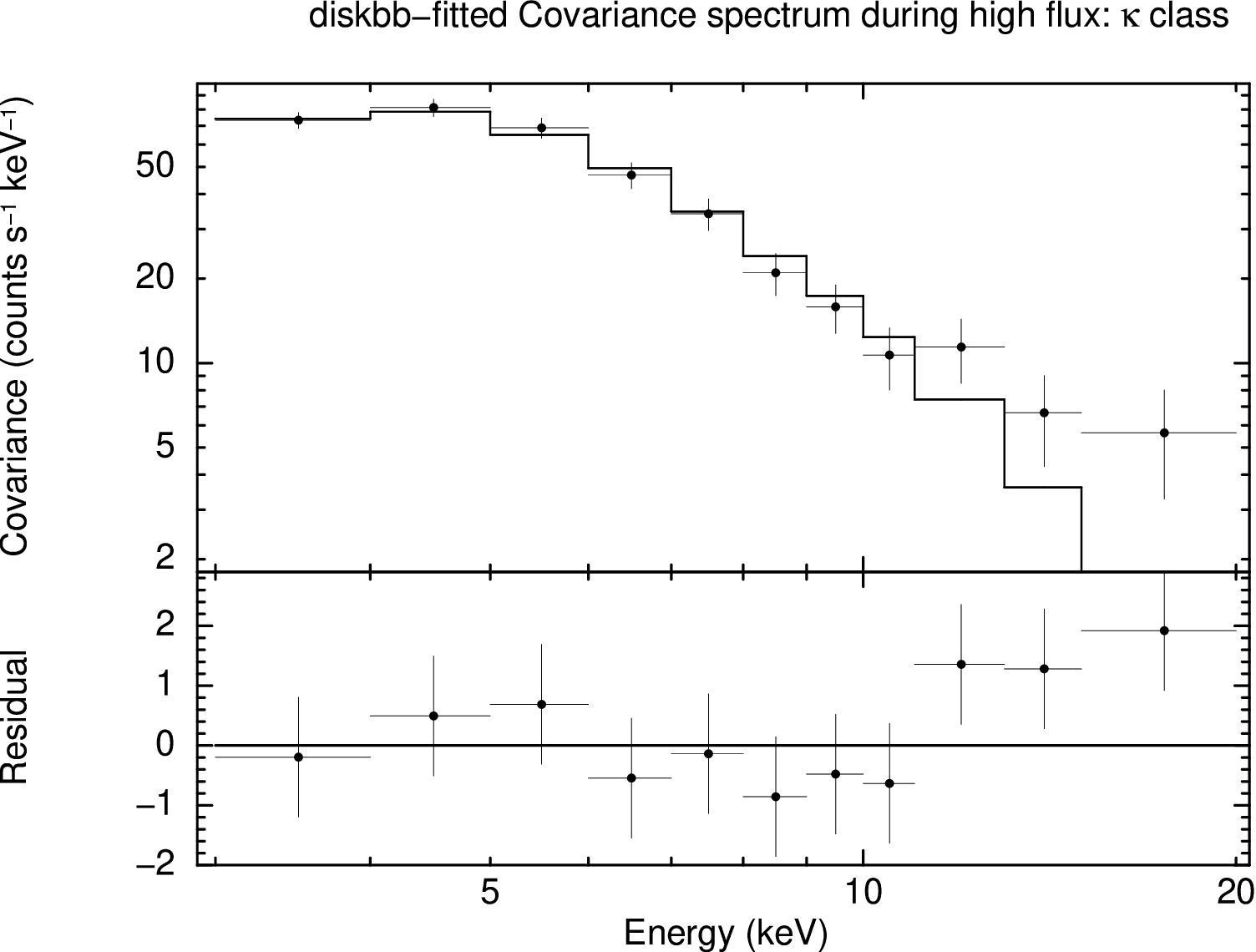}
    \hfill
    \includegraphics[width=0.35\textwidth,angle=270]{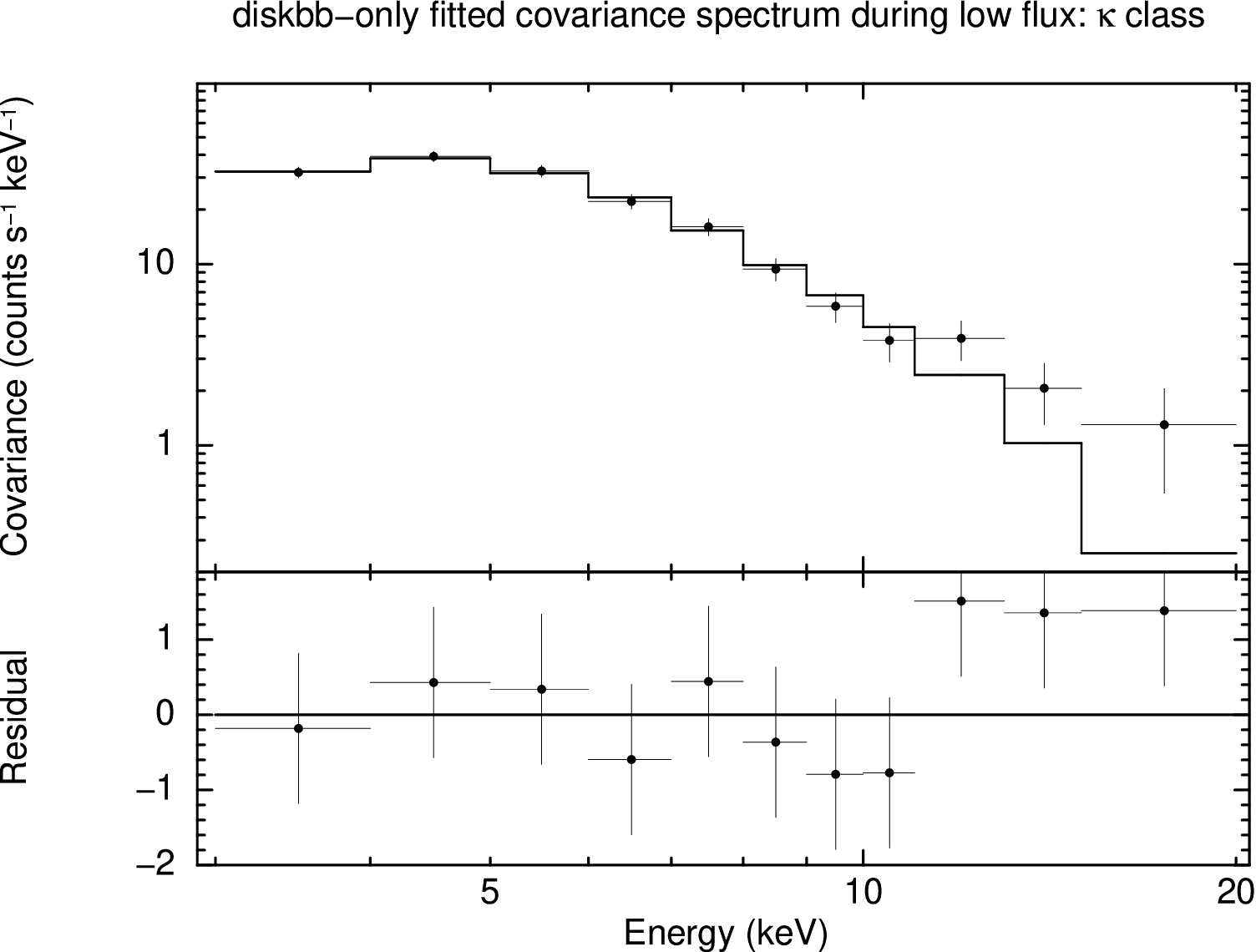}
    \hfill
    \includegraphics[width=0.35\textwidth,angle=270]{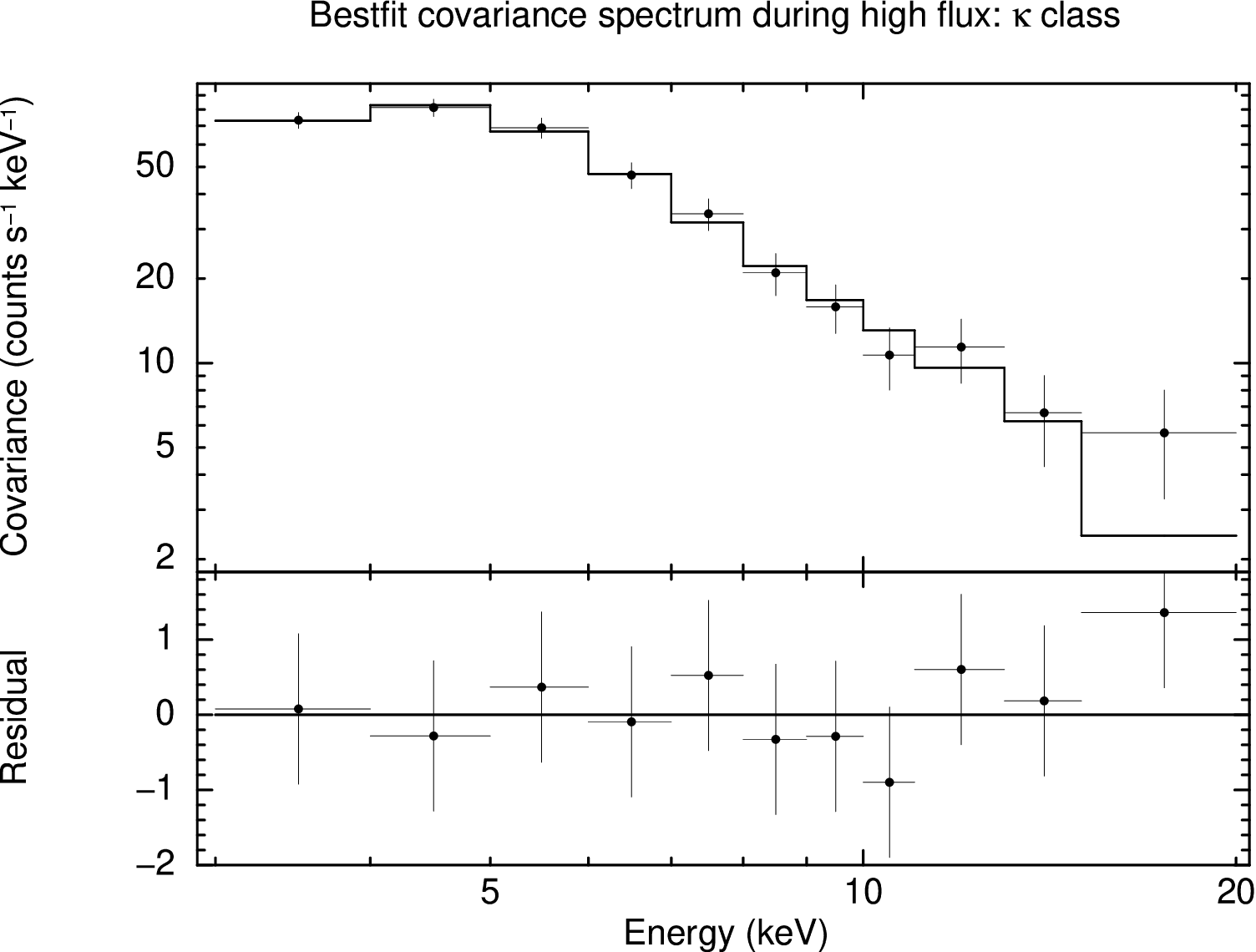}
    \hfill
    \includegraphics[width=0.35\textwidth,angle=270]{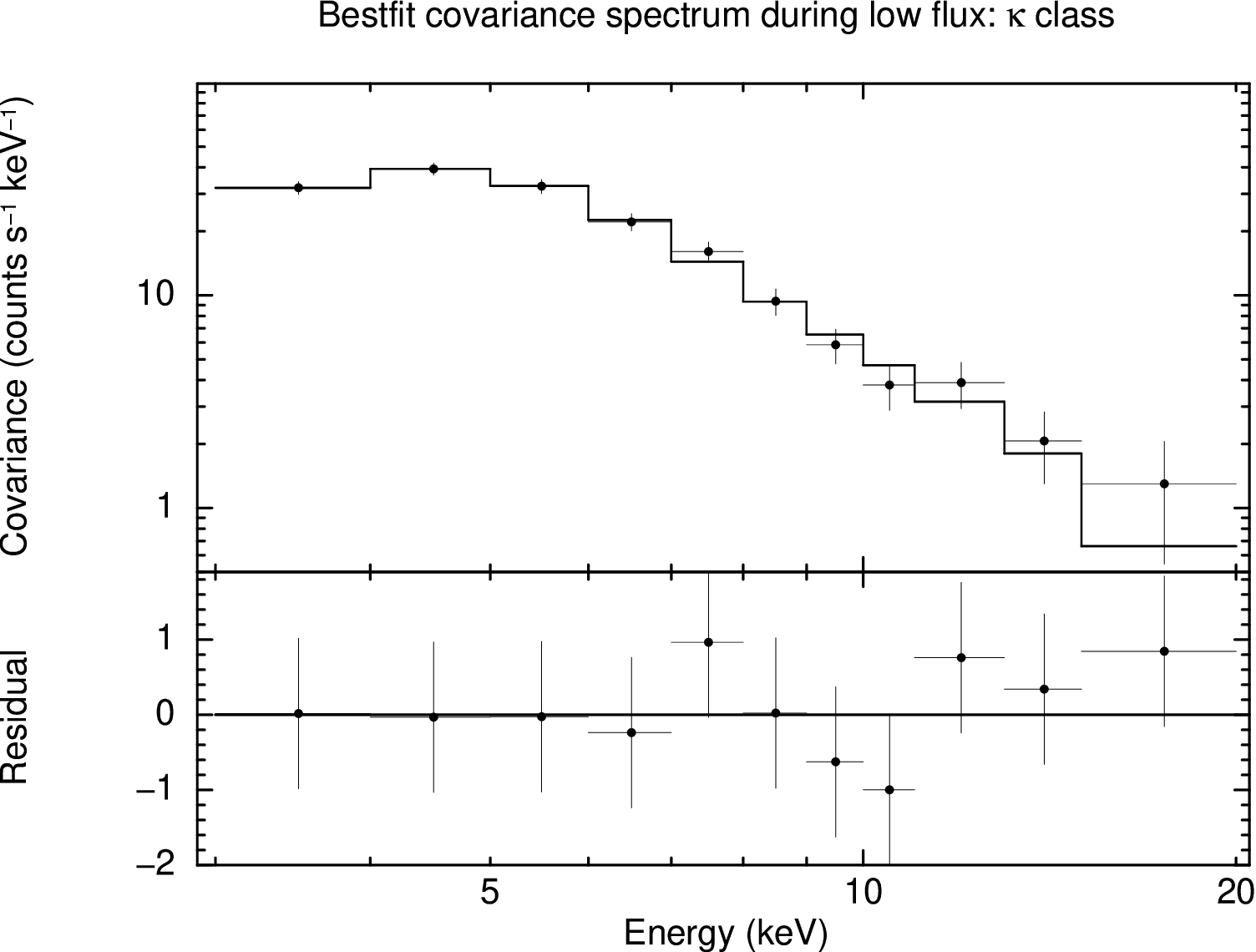}
    \caption{Modelling of the covariance spectra for $\kappa$ class in the Fourier frequency range 0.01-5.0 Hz  are shown with the model components and fit residual. The top-left panel presents the spectral fits using only a diskbb component for the high-flux state, while the top-right panel shows similar spectral fits for the low-flux state. The bottom-left panel displays the best-fit covariance spectra, modeled with a combination of diskbb and Comptonized components for the high-flux state, and the bottom-right panel shows the same model applied to the low-flux state.}
    \label{fig4}
\end{figure*}

\begin{figure*}[!ht]
    \centering
    \includegraphics[width=0.35\textwidth,angle=270]{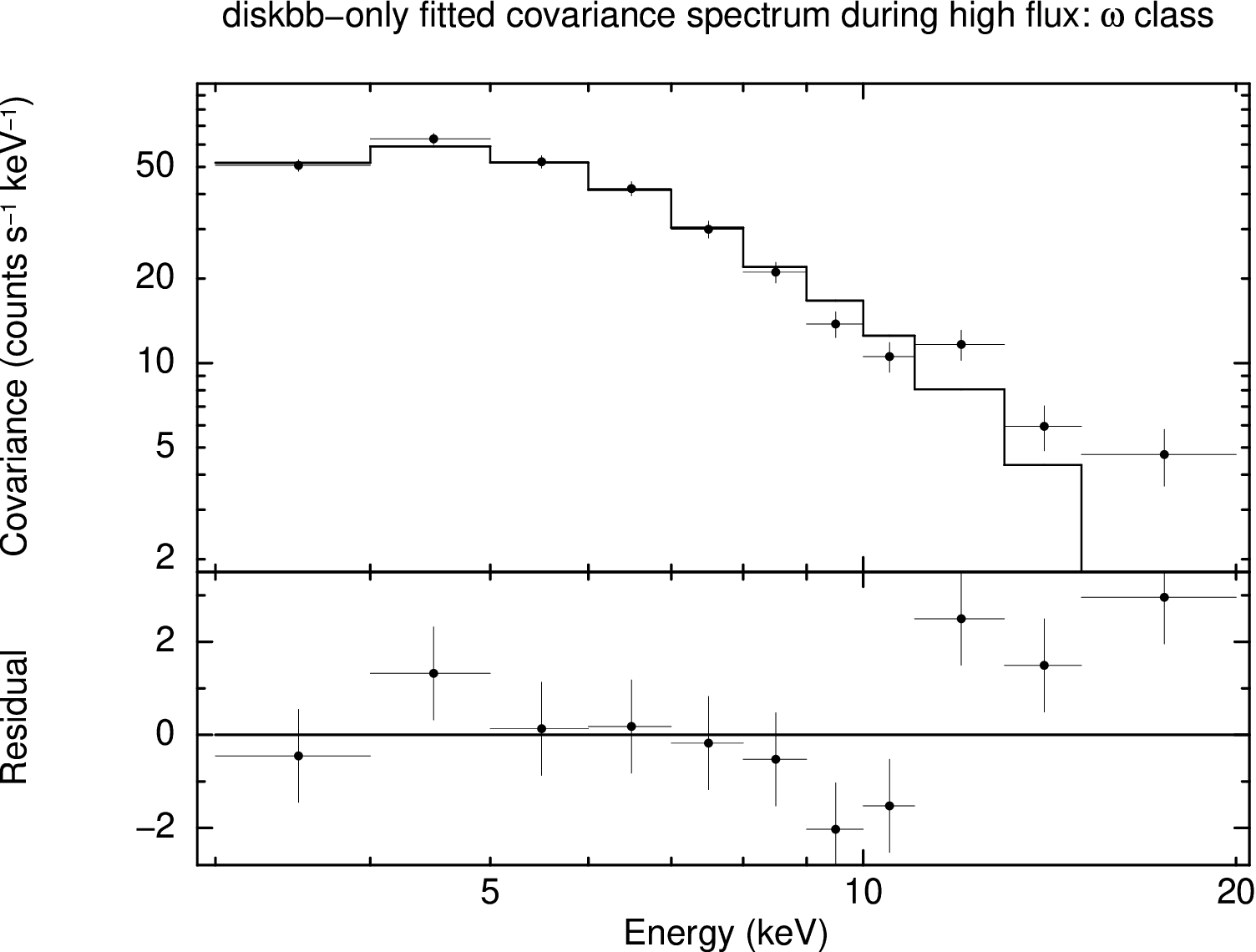}
    \includegraphics[width=0.35\textwidth,angle=270]{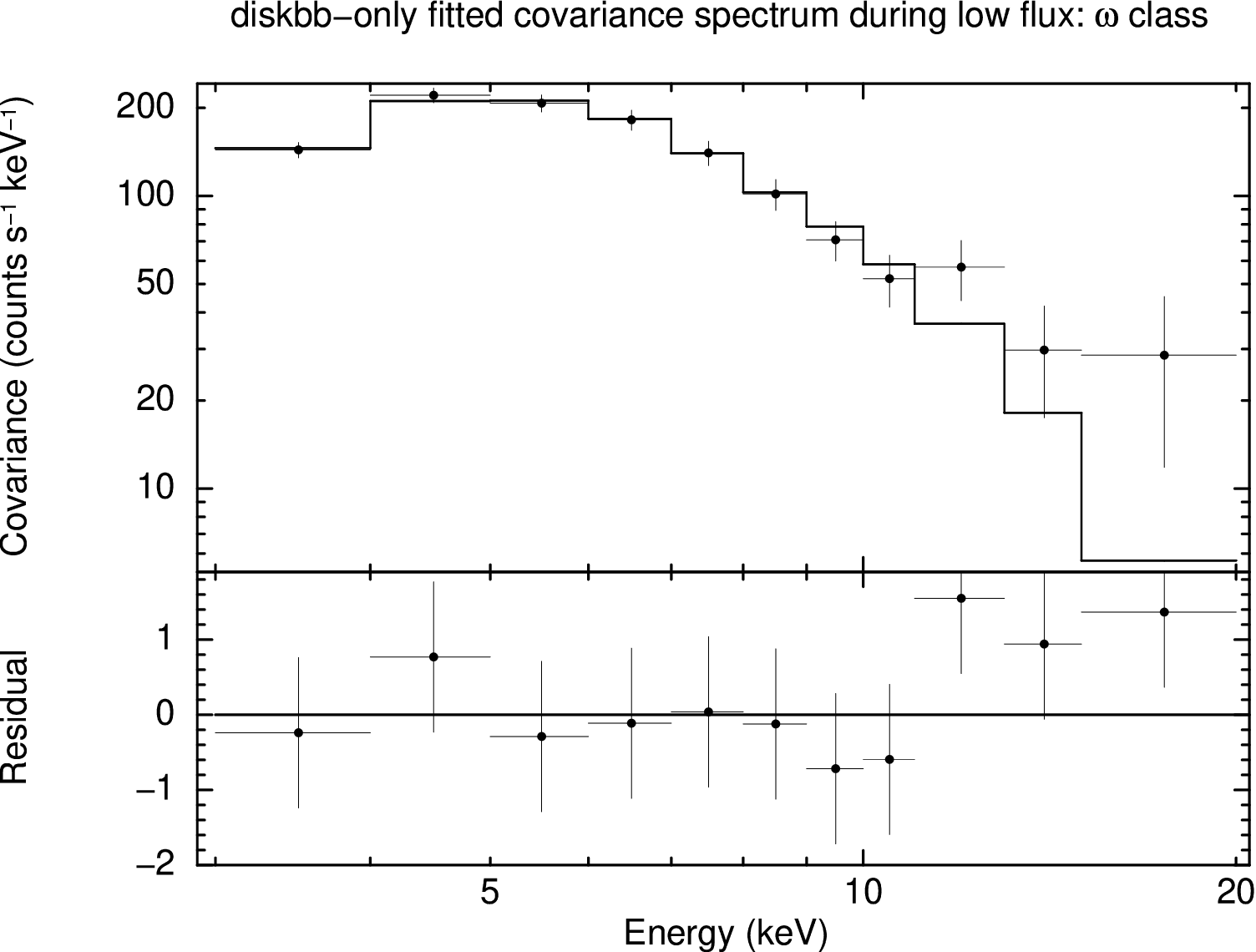}
    \includegraphics[width=0.35\textwidth,angle=270]{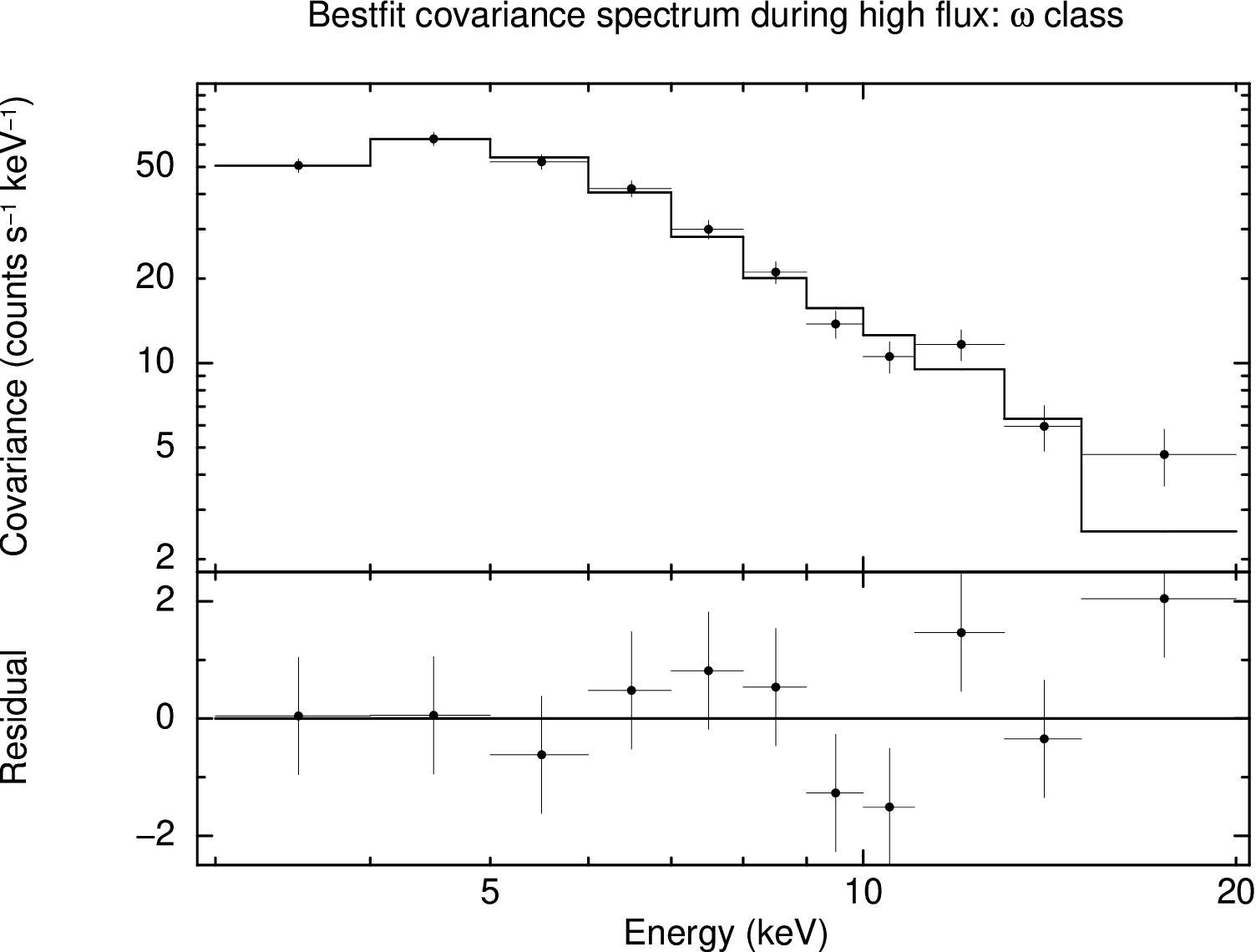}
    \includegraphics[width=0.35\textwidth,angle=270]{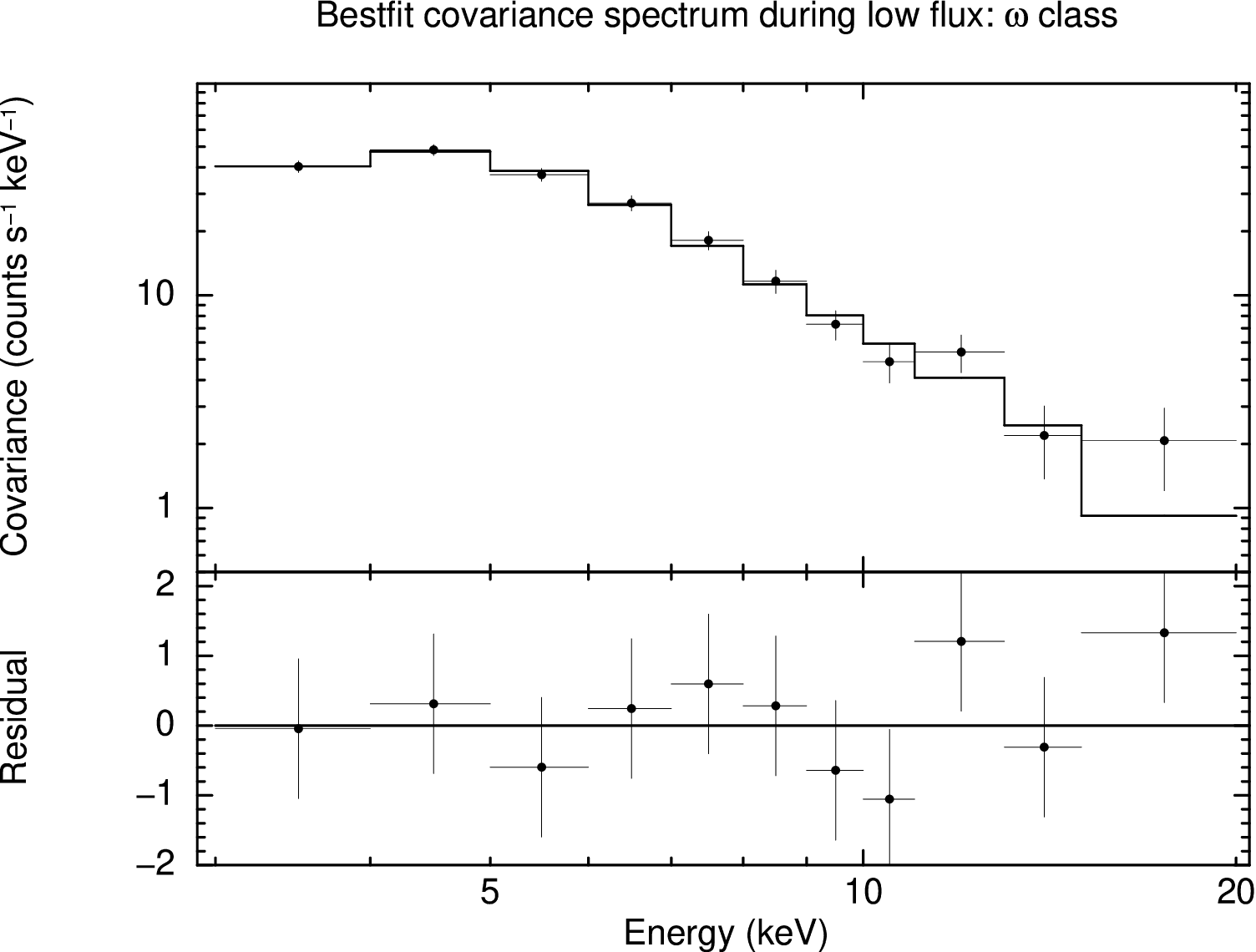}
    \caption{Modeling of the covariance spectra in the 0.01-5.0 Hz Fourier range, with the best-fit covariance energy spectra for the $\omega$ class shown alongside the model components and residuals. The top-left panel presents the spectral fits using only a diskbb component for the high-flux state, while the top-right panel shows similar spectral fits for the low-flux state. The bottom-left panel displays the best-fit covariance spectra, modeled with a combination of diskbb and Comptonized components for the high-flux state, and the bottom-right panel shows the same model applied to the low-flux state.}
    \label{fig5}
\end{figure*}

For each flux interval, we fitted the covariance spectra using simplified disc-plus-Compton models in order to determine whether the rapidly varying emission is dominated by the thermal disc, the Comptonized continuum, or a mixture of both. The best-fitting models are presented in Figs \ref{fig4} and \ref{fig5}, and the corresponding parameter values are listed in Tables \ref{tab5} and \ref{tab6}. The covariance spectra generally exhibit smooth, curved shapes that rise toward lower energies, strongly suggesting that the bulk of the coherent variability on sub-second timescales is associated with changes in the thermal disc emission. In contrast, the Comptonized tail contributes only weakly to the covariance, and in several cases it appears to be 
consistent with little or no coherent variability in the frequency range probed. This distinction already indicates that the corona does not respond as rapidly or as coherently as the disc to the driving fluctuations.

\subsection{Covariance Spectra of the $\kappa$ Class}
The covariance spectra of the $\kappa$ class reveal a clear difference between the high- and low-flux intervals. In the high-flux interval, the covariance is strongly peaked at the lowest energies and decreases steadily towards 20\,keV. When fitted with a disc-dominated model, the shape corresponds to a thermal spectrum with an effective temperature of approximately 2.3 keV. The local absorption column density required to model the high-flux covariance is somewhat higher than that obtained from the time-averaged spectra, indicating that the variable emission may arise preferentially from regions with additional obscuration or enhanced local absorption.

During the low-flux interval, the covariance spectrum retains the same overall curvature but peaks at slightly lower energies. The best-fitting disc temperature decreases to values near 1.8 keV, indicating that the variability originates from a cooler disc during the dip. The overall amplitude of the covariance also declines, reflecting the substantial reduction in soft-band flux. The Comptonized component remains only weakly required by the fits, as the inclusion of a power-law improves the residuals only marginally and yields photon indices that are not well constrained. This behaviour confirms that in the $\kappa$ class, the fast variations are dominated by temperature changes in the disc rather than by variations in the corona.

\subsection{Covariance Spectra of the $\omega$ Class}

The covariance spectra of the $\omega$ class display similar characteristics but with some distinctive differences in detail. In the high-flux interval, the covariance spectrum is again dominated by a soft component that is well described by a disc blackbody model, although the effective temperature inferred from the fits is lower than in the $\kappa$ class, falling between 1.2 and 1.4 keV. This suggests that even in the bright phase, the rapidly varying emission of the $\omega$ class originates from a comparatively cooler disc. The corresponding absorption column density is again higher than in the mean-spectrum fits, a result that is consistent with the idea that the variable disc emission arises close to the innermost regions, where dense inhomogeneities or partial covering may be present.
In the low-flux interval of the $\omega$ class, the covariance spectra change shape in a manner similar to that observed in the $\kappa$ class. The peak shifts to lower energies, and the best-fitting disc temperature increases slightly from that of the high-flux covariance, lying between 2.0 and 2.2\,keV across the two observations. This counterintuitive behaviour, in which the covariance spectrum appears hotter in the low-flux interval, arises because the fractional variability is dominated by the component of the disc that continues to fluctuate coherently even as the overall disc emission cools and recedes. The amplitude of the covariance decreases substantially, matching the reduction in total flux seen during the low-flux intervals.
The fits again show little requirement for a Comptonized contribution, and in several cases a pure disc model provides the best statistical description of the covariance. Only in the high-flux interval of Obs.~$\omega_2$ do we obtain marginal evidence for a variable power-law tail, but the improvement is not statistically significant. The absence of strong variability in the Comptonized component suggests that the corona responds on longer timescales than those probed in our Fourier frequency range, or that its intrinsic fluctuations are less coherent with those of the disc.

\subsection{Interpretation of Covariance Behaviour}
A comparison of the covariance spectra across all observations indicates that the rapid variability in both the $\kappa$ and $\omega$ classes is driven predominantly by the accretion disc. The temperatures inferred from the covariance spectra differ from those obtained from the time-averaged spectra, implying that the subset of the disc responsible for the coherent fluctuations is not identical to the disc component that dominates the total flux. This distinction is consistent with a scenario in which fluctuations propagate inward through the disc and preferentially modulate the innermost radii, resulting in hotter or more coherently varying regions than those inferred from the mean spectrum.
The near-absence of rapidly varying Comptonized emission indicates that the corona either responds more slowly to the driving fluctuations or that its variability is dominated by incoherent processes that do not correlate strongly with the disc on sub-second timescales. This finding complements the results of the time-averaged spectral analysis, which shows that the corona becomes harder and energetically more dominant during the low-flux intervals, but does not vary coherently on the short timescales responsible for the flux oscillations.
Taken together, the covariance results demonstrate that the fastest variability in both the $\kappa$ and $\omega$ classes originates primarily in the accretion disc and is associated with changes in disc temperature or normalisation on timescales of a few tenths of a second. The coronal component, although it hardens significantly during the flux dips, contributes comparatively little to the coherent rapid variability. These findings provide strong evidence that the disc--corona system responds differently to the underlying instability that drives the limit-cycle behaviour, with the disc undergoing rapid coherent oscillations while the corona evolves more gradually or stochastically. The covariance analysis, therefore, reinforces the interpretation that the $\kappa$ and $\omega$ classes arise from repeated cycles of disc evacuation and refilling driven by radiation-pressure instability, with the coronally emitted hard flux playing a secondary role in the fast coherent variability.
\begin{table}[!ht]
\centering
\caption{Best-fit covariance spectral parameters for flux-resolved $\kappa_1$ and $\kappa_2$ class observations during high and low flux intervals. Bestfit model: TBabs*diskbb}
\begin{tabular}{lcccccc}
\hline
Parameter & Units & $\kappa_1$ (High) & $\kappa_1$ (Low) & $\kappa_2$ (High) & $\kappa_2$ (Low) \\
\hline
$N_{\rm H}$ & $10^{22}\,\mathrm{cm}^{-2}$ & $7.48^{+1.09}_{-1.16}$ & $10.31^{+1.47}_{-1.43}$ & $9.43^{+1.63}_{-1.81}$ & $11.25^{+1.24}_{-1.57}$ \\
Disc temperature ($kT_{\rm in}$) & keV & $2.34^{+0.22}_{-0.19}$ & $1.82^{+0.19}_{-0.22}$ & $2.19^{+0.21}_{-0.17}$ & $1.31^{+0.24}_{-0.21}$ \\
Disc Norm (x 10$^5$) &  & $11.9^{+6.3}_{-4.1}$ & $12.9^{+5.4}_{-3.8}$ & $10.8^{+5.5}_{-4.6}$ & $13.1^{+20.2}_{-7.6}$ \\
$\chi^2$/dof & & 9/8 & 8/8 & 8/8 & 7/8 \\
\hline
\end{tabular}
\label{tab5}
\end{table}
\begin{table}[!ht]
\centering
\caption{Best-fit covariance spectral parameters for $\omega$ class observations in high and low flux intervals. Bestfit model: TBabs*(THcomp*diskbb)}
\begin{tabular}{lcccccc}
\hline
\hline
Parameter & Units & $\omega_1$ (High) & $\omega_1$ (Low) & $\omega_2$ (High) & $\omega_2$ (Low) \\
\hline
$N_{\rm H}$ & $10^{22}\,\mathrm{cm}^{-2}$ & $10.22^{+1.39}_{-1.22}$ & $10.32^{+1.92}_{-1.81}$ & $7.87^{+1.23}_{-1.22}$ & $10.23^{+1.91}_{-1.76}$ \\
Disc temperature ($kT_{\rm in}$) & keV & $1.36^{+0.16}_{-0.14}$ & $2.01^{+0.12}_{-0.11}$ & $1.33^{+0.24}_{-0.19}$ & $2.16^{+0.13}_{-0.12}$ \\
Disc Norm (x 10$^5$)&  & $13^{+11}_{-6}$ & $14.7^{+5.8}_{-3.4}$ & $13.9^{+20.8}_{-8.2}$ & $10.5^{+3.6}_{-2.7}$ \\
Power-law index ($\Gamma$) &  & $1.64^{+0.27}_{-0.20}$ & -- & $1.79^{+0.28}_{-0.18}$ & -- \\
$\chi^2$/dof & & 11/7 & 7/8 & 9/7 & 10/8 \\
$\chi^2$/dof (without thcomp) & & 18/8 & -- & 17/8 & -- \\
\hline
\end{tabular}
\label{tab6}
\end{table}

\section{Discussion and Conclusion}\label{sec6}
\label{sect:discussion}
In this work, we present a comprehensive spectro–temporal study of the $\kappa$ and $\omega$ variability classes of GRS~1915+105 using \textit{AstroSat} SXT and LAXPC observations. The lightcurves reveal large-amplitude, quasi-periodic oscillations with sharp flux transitions occurring on timescales of a few tens of seconds. Flux-resolved spectral analysis shows that the high-flux intervals correspond to a hot, disc-dominated state with $kT_{\rm in}\approx4$~keV and a steep Comptonized tail, whereas the low-flux intervals exhibit a significantly cooler or partially recessed disc with $kT_{\rm in}\approx2.3$--$2.7$~keV and a harder coronal spectrum, indicating reduced soft-photon cooling. Covariance spectra demonstrate that the fast, coherent variability (0.015–5~Hz) arises almost entirely from the disc, while the corona contributes little on these timescales. Using standard thin-disc scaling, the observed temperature drop implies an increase in the inner disc radius by a factor of $\sim$1.8-2, which yields viscous timescales consistent with the observed recurrence periods for a $12.4\,M_{\odot}$ black hole. Together, these results support a unified picture in which radiation-pressure driven limit-cycle oscillations govern the repeated disc evacuation and refilling that produce the characteristic burst–dip behaviour of the $\kappa$ and $\omega$ classes.

Our results strongly reinforce and extend the limit-cycle interpretation originally proposed by \citet{Belloni1997}. The observed high-flux intervals in both the $\kappa$ and $\omega$ classes correspond to hot, optically thick inner discs accompanied by steep Comptonized emission, whereas the low-flux intervals are characterised by a cooler or partially recessed disc and a harder coronal continuum. The systematic decrease in disc temperature by $\sim$1--2~keV and the simultaneous hardening of the Comptonized component during the dips are in excellent agreement with the cyclic disc evacuation and refilling scenario envisioned by \citet{Belloni1997}.
 
The principal advance of the present work is the inclusion of covariance spectral diagnostics, which demonstrate that the rapid ($0.01$--$5$~Hz) coherent variability is dominated almost entirely by the thermal disc, while the Comptonized component contributes only weakly on these timescales. This provides direct spectro-timing confirmation of the qualitative picture inferred by \citet{Belloni1997}, namely that the accretion disc is the primary driver of the oscillations, with the corona responding more slowly through changes in cooling efficiency rather than through coherent intrinsic variability.
 
Thus, rather than introducing a new physical mechanism, our findings place the $\kappa$ and $\omega$ variability classes on a firmer quantitative footing within the \citet{Belloni1997} framework, demonstrating that radiation-pressure-driven inner disc instabilities remain the most natural explanation for the observed large-amplitude oscillations in GRS~1915+105.

To quantify whether the observed temperature change is consistent with the characteristic timescales expected for a $\sim 12.4 M_{\odot}$ black hole, we start from the standard thin-disc scaling for the effective temperature near the inner disc edge.  If the dominant changes are geometric (i.e. changes in the
apparent inner radius $R_{\rm in}$) Rather than large changes in the global accretion rate, the disc temperature at the inner edge scales approximately as
\begin{equation}
    T_{\rm in}\ \propto\ \dot{M}^{1/4}\,M^{-1/2}\,r_{\rm in}^{-3/4},
    \label{eq:Tscaling}
\end{equation}
where $r_{\rm in}=R_{\rm in}/R_{\rm g}$ is the inner radius in gravitational units and $R_{\rm g}=GM/c^{2}$.
For fixed mass $M$ and approximately constant short-term mass supply $\dot{M}$, the ratio of inner radii
between high and low states follows directly from the measured temperatures:
\begin{equation}
    \frac{r_{\rm in,low}}{r_{\rm in,high}} \;=\; \left(\frac{T_{\rm in,high}}{T_{\rm in,low}}\right)^{4/3}.
    \label{eq:rratio}
\end{equation}
Using representative values from our fits (for example $T_{\rm in,high}=4.0\,$keV and
$T_{\rm in,low}=2.5\,$keV) gives
\[
    \frac{r_{\rm in,low}}{r_{\rm in,high}} \;=\; \left(\frac{4.0}{2.5}\right)^{4/3} \approx 1.87,
\]
so that an inner radius initially near $r_{\rm in,high}\sim 6\,R_{\rm g}$ (close to the ISCO for modest spin) would move out to $r_{\rm in,low}\sim 11.2\,R_{\rm g}$ in the low state.  Such a factor of $\sim$1.8--2 increase in the apparent inner radius is consistent in magnitude with the spectral evolution we measure. The physical mechanism driving repeated evacuation and refilling of the inner disc on the observed timescales are expected to be a radiation-pressure–dominated thermal/viscous instability \citep{Neilsen2011}. The natural timescale for limit-cycle behaviour is of order the viscous timescale in the radiation-pressure dominated region. A useful approximation for the local viscous time is
\begin{equation}
    t_{\rm visc}(R)\ \simeq\ \frac{1}{\alpha}\left(\frac{R}{H}\right)^{2} t_{\rm orb}(R),
    \label{eq:tvisc}
\end{equation}
where $\alpha$ is the Shakura--Sunyaev viscosity parameter, $H$ is the local scale height, and
\begin{equation}
    t_{\rm orb}(R) \;=\; 2\pi \sqrt{\frac{R^{3}}{GM}}
    \label{eq:torb}
\end{equation}
is the orbital period at radius $R$.  Writing $R=r\,R_{\rm g}$ and substituting $R_{\rm g}=GM/c^{2}$ gives
\begin{equation}
    t_{\rm orb}(r)\;=\;2\pi\,r^{3/2}\,\frac{GM}{c^{3}}.
    \label{eq:torb_rg}
\end{equation}
For the black-hole mass adopted in this work, $M=12.4\,M_{\odot}$, one finds numerically
\[
    \frac{GM}{c^{3}} \approx 6.17\times 10^{-6}\ {\rm s},
\]
so that the orbital period at $r\simeq 11.2$ (the post-transition radius estimated above) is
\[
    t_{\rm orb}(r\!=\!11.2)\;\approx\;2\pi\,(11.2)^{3/2}\,(6.17\times10^{-6}{\rm s}) \approx 0.0144\ {\rm s}.
\]
The viscous time at that radius then depends sensitively on the uncertain parameters $\alpha$ and $H/R$. For a plausible range of parameters expected in near-Eddington, radiation-pressure dominated inner discs, taking $\alpha$ in the range $0.01$--$0.1$ and $H/R$ in the range $0.05$--$0.2$, equation (\ref{eq:tvisc}) yields viscous timescales at $r\sim 11$ of order a few seconds up to several hundred seconds. Evaluating representative combinations gives, for example,
\[
\begin{array}{l l}
H/R=0.20,\ \alpha=0.10: & t_{\rm visc}\approx 3.6\ {\rm s},\\[2pt]
H/R=0.10,\ \alpha=0.10: & t_{\rm visc}\approx 14.4\ {\rm s},\\[2pt]
H/R=0.07,\ \alpha=0.10: & t_{\rm visc}\approx 29.5\ {\rm s},\\[2pt]
H/R=0.05,\ \alpha=0.10: & t_{\rm visc}\approx 57.8\ {\rm s}.
\end{array}
\]
If the effective viscosity is lower ($\alpha\sim 0.01$) these numbers increase by an order of magnitude, overlapping the upper end of the observed recurrence times (tens to hundreds). Given the uncertainties in the vertical structure and the local $H/R$ in a radiation-pressure dominated flow, these estimated viscous times are entirely compatible with the observed burst recurrence times of $\sim$20--100\,s
for the $\kappa$ and $\omega$ classes.  In short, the magnitude of the measured temperature drop, interpreted as an increase in $R_{\rm in}$ by a factor $\sim$1.8--2, naturally leads to viscous timescales at the new inner radius that are consistent with the observed oscillation periods for physically plausible values of $\alpha$ and $H/R$.

The covariance results are consistent with this interpretation. The covariance spectra attribute the fastest coherent variability to the thermal disc component, indicating that the part of the disc that fluctuates coherently is located at small radii and therefore responds on short local thermal/viscous timescales. The corona, by contrast, does not display strong coherent variability in the 0.015--5 Hz band, consistent with the idea that the corona's state is set more slowly by the changing soft-photon supply as the disc evacuates and refills. Thus, the combined spectro-temporal evidence favours a limit-cycle model in which thermal/viscous processes in a radiation-pressure dominated inner disc drive the high--to--low transitions, while the coronal hardening is a reactive effect arising from reduced Compton cooling.

Our results are broadly consistent with recent AstroSat-based studies that link variability classes and HFQPO behaviour in GRS~1915+105 to changes in the inner accretion flow \citep{Majumder2022, Dhaka2025, Harikesh2025}. These works demonstrate that HFQPO properties evolve systematically across flaring and oscillatory classes, implying that the geometry and physical conditions of the innermost flow region vary on short timescales. Within this broader context, our findings provide a complementary perspective by showing that the large-amplitude flux oscillations in the $\kappa$ and $\omega$ classes are primarily driven by rapid structural changes in the thermal accretion disc.
 
In particular, the covariance spectral analysis reveals that the fast ($0.01$--$5$~Hz) coherent variability arises almost entirely from the disc, while the Comptonized component contributes weakly on these timescales. This suggests that the disc evacuation and refilling cycles inferred from our flux-resolved spectroscopy likely set the boundary conditions for the inner-flow dynamics probed by HFQPOs. In this picture, HFQPOs may originate in the innermost regions of the flow whose characteristic frequencies respond to the evolving disc truncation radius and local physical conditions during the limit-cycle oscillations. Our results, therefore, provide a physical framework within which the HFQPO phenomenology reported by \citet{Majumder2022}, \citet{Dhaka2025}, and \citet{Harikesh2025} can be naturally interpreted.

We emphasise several caveats.  First, the mapping from measured colour temperature to the effective emitting radius is affected by spectral hardening (colour correction) and any additional scattering or partial-covering absorption; corrections of order unity to $r_{\rm in}$ are therefore possible. Second, the vertical structure ($H/R$) and the effective $\alpha$ in the radiation-pressure regime are not well constrained observationally and may depend on magnetic stresses and wind/jet coupling.  Nevertheless, the simple, quantitative estimate presented above shows that the observed magnitude of the disc cooling and implied change in inner radius are compatible with viscous timescales of the right order for a $\sim 12.4 M_{\odot}$ black hole, and therefore support the interpretation that radiation-pressure driven
limit-cycle oscillations underlie the $\kappa$ and $\omega$ phenomenology.

\normalem
\begin{acknowledgements}
We thank the referee for constructive suggestions and comments which help to improve the manuscript. The present work makes use of data from the AstroSat mission of the Indian Space Research Organisation (ISRO), archived at the Indian Space Science Data Centre (ISSDC). This work has been performed utilizing the calibration databases and auxiliary analysis tools developed, maintained, and distributed by AstroSat/LAXPC teams with members from various institutions in India and abroad. This work has used data from the Soft X-ray Telescope (SXT), developed at TIFR, Mumbai, and the SXT POC at TIFR is thanked for verifying and releasing the data via the ISSDC data archive and providing the necessary software tools. The research work at the Indian Institute of Technology, Hyderabad, is financially supported or funded by the Council of Scientific and Industrial Research (CSIR-Grant No: 09/1001(12694)/2021-EMR-I), Ministry of Science and Technology, Government of India.
\end{acknowledgements}


\newpage
\section{Appendix}
\begin{figure*}[!ht]
    \centering
    \includegraphics[width=0.37\textwidth,angle=270]{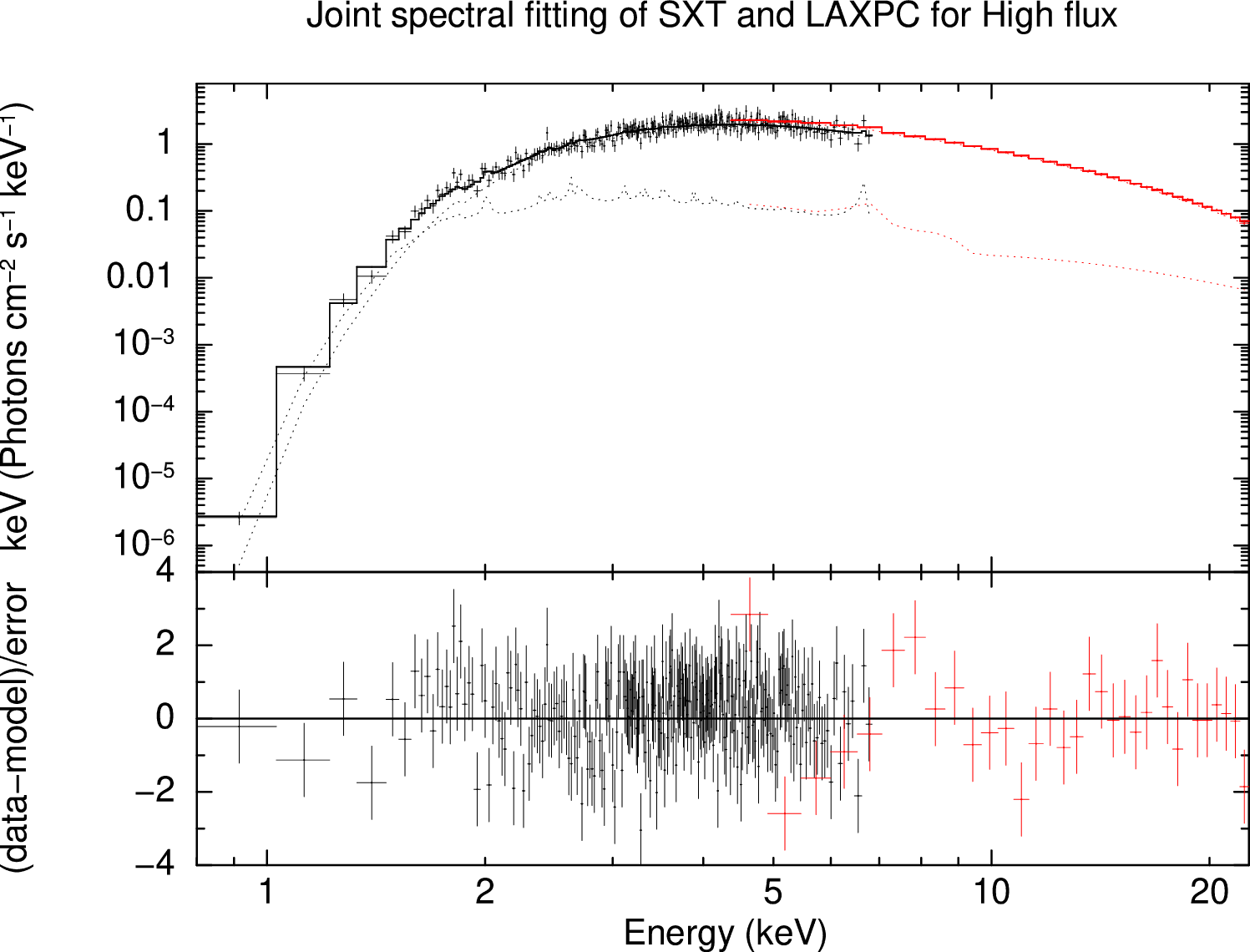}
    \includegraphics[width=0.37\textwidth,angle=270]{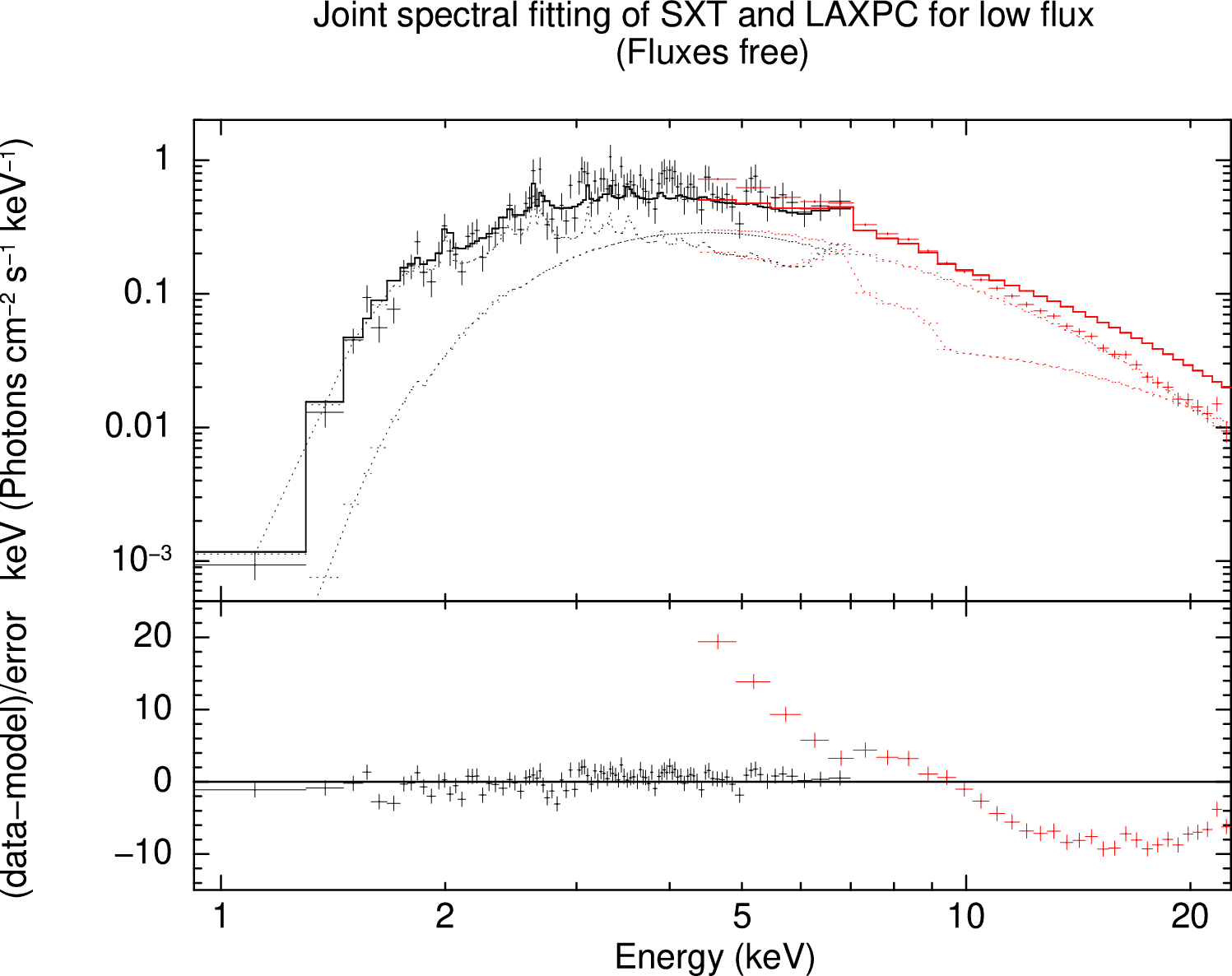}
    \includegraphics[width=0.37\textwidth,angle=270]{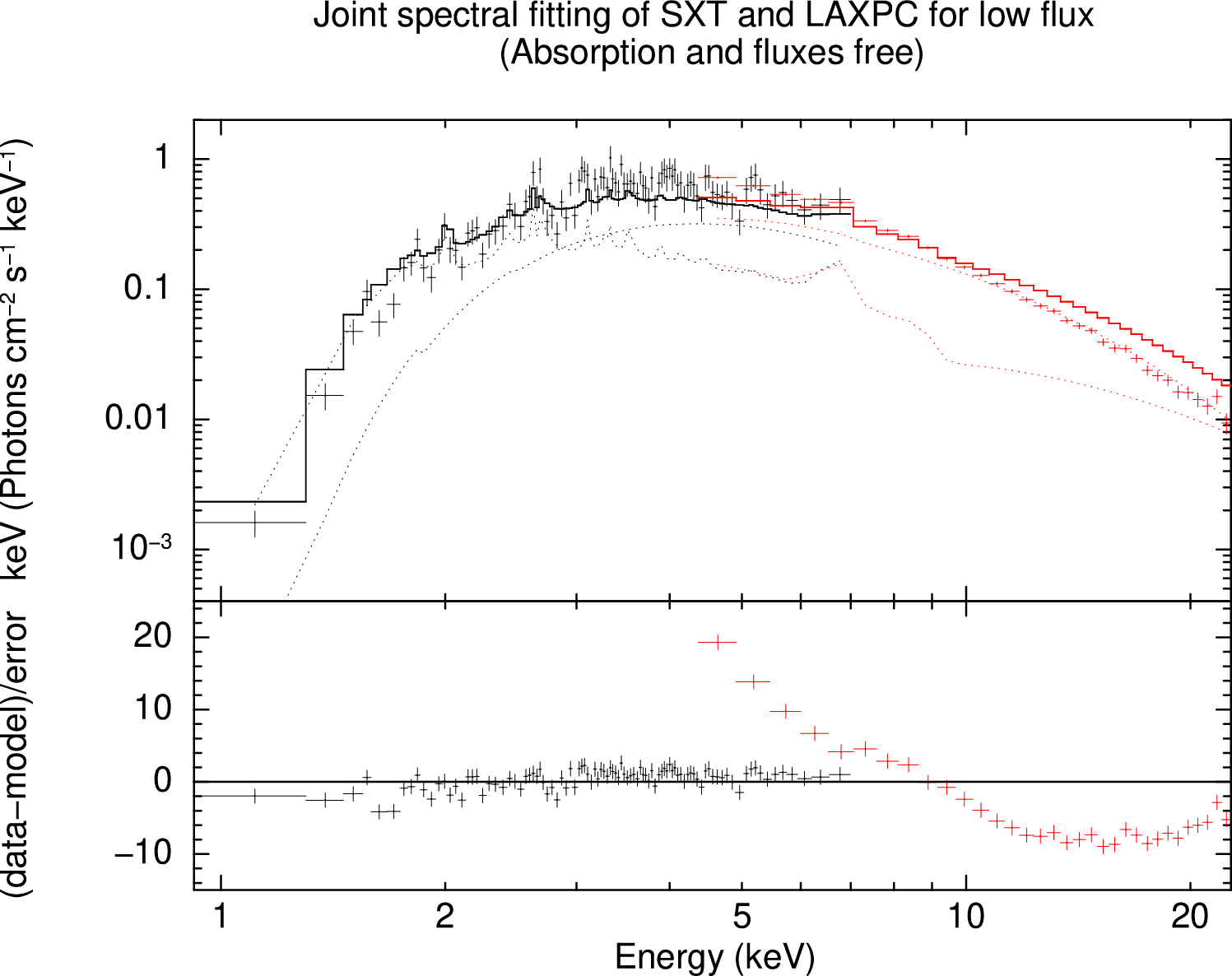}
    \includegraphics[width=0.37\textwidth,angle=270]{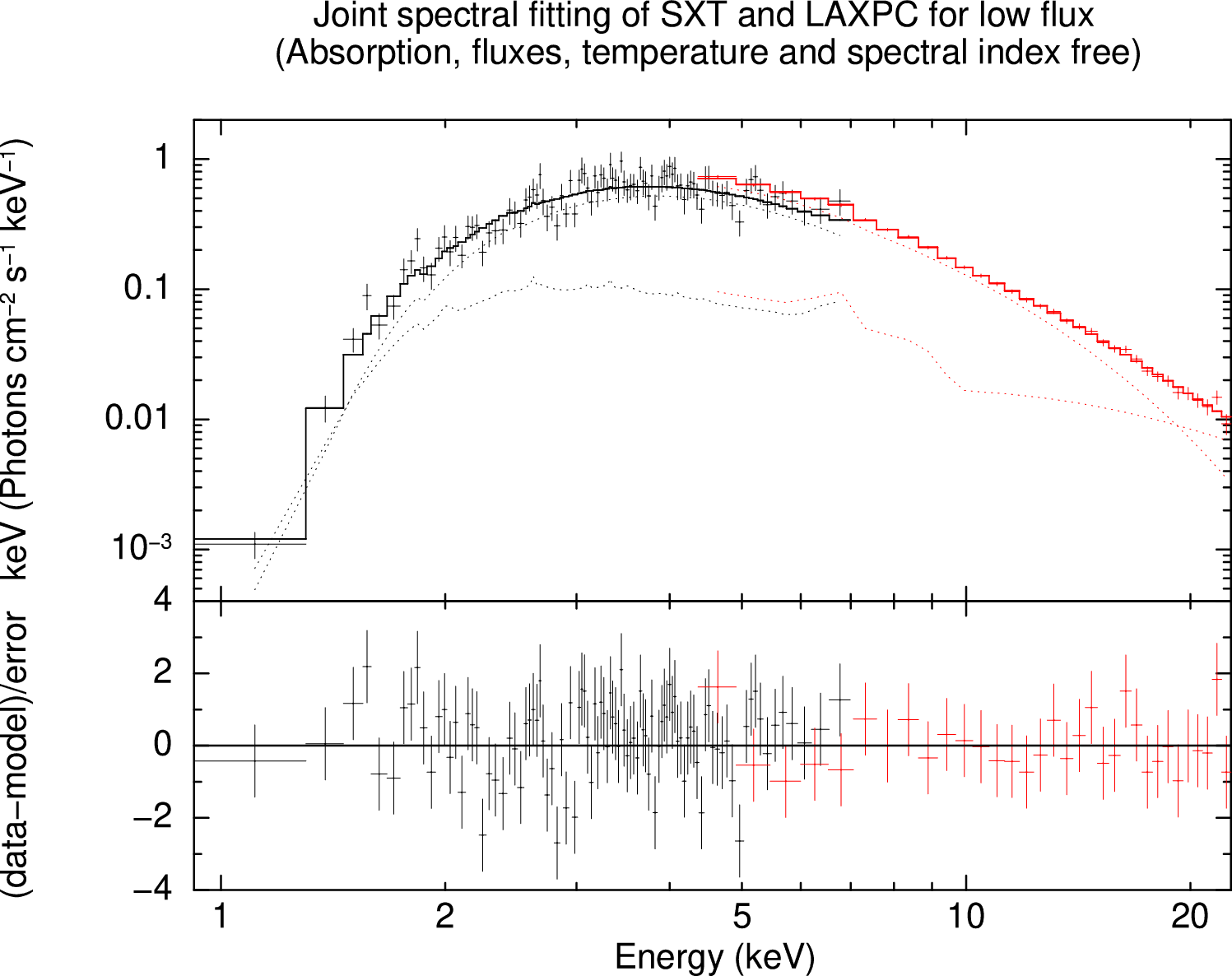}
    \caption{Unfolded \asat{} SXT/LAXPC20 mean spectra of \src{} in the 1.0–25.0 keV energy range. The upper-left panel shows the best-fit joint spectra for the high-flux state of the $\kappa$ class (Obs. $\kappa_1$), including the model components and data-to-model ratio. The upper-right, and lower-left panels display the joint spectral fits for Case-A, and Case-B, respectively, corresponding to the low-flux state of the $\kappa$ class (Obs. $\kappa_1$). The lower-right panel presents the best-fit joint spectra for the low-flux (Case-C) of the $\kappa$ class.}
    \label{fig7}
\end{figure*}
\begin{figure*}[!ht]
    \centering
    \includegraphics[width=0.37\textwidth,angle=270]{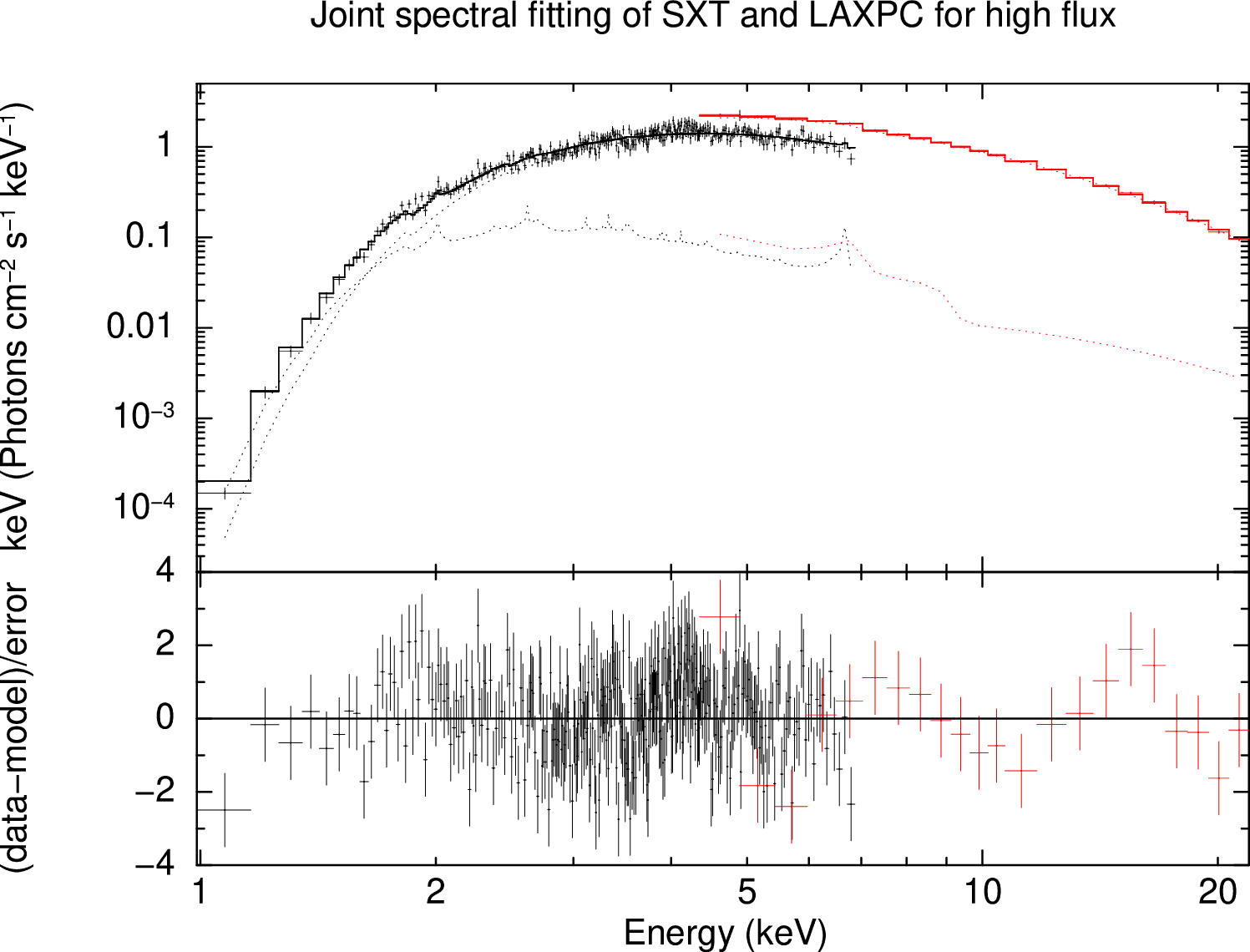}
    \hfill
    \includegraphics[width=0.37\textwidth,angle=270]{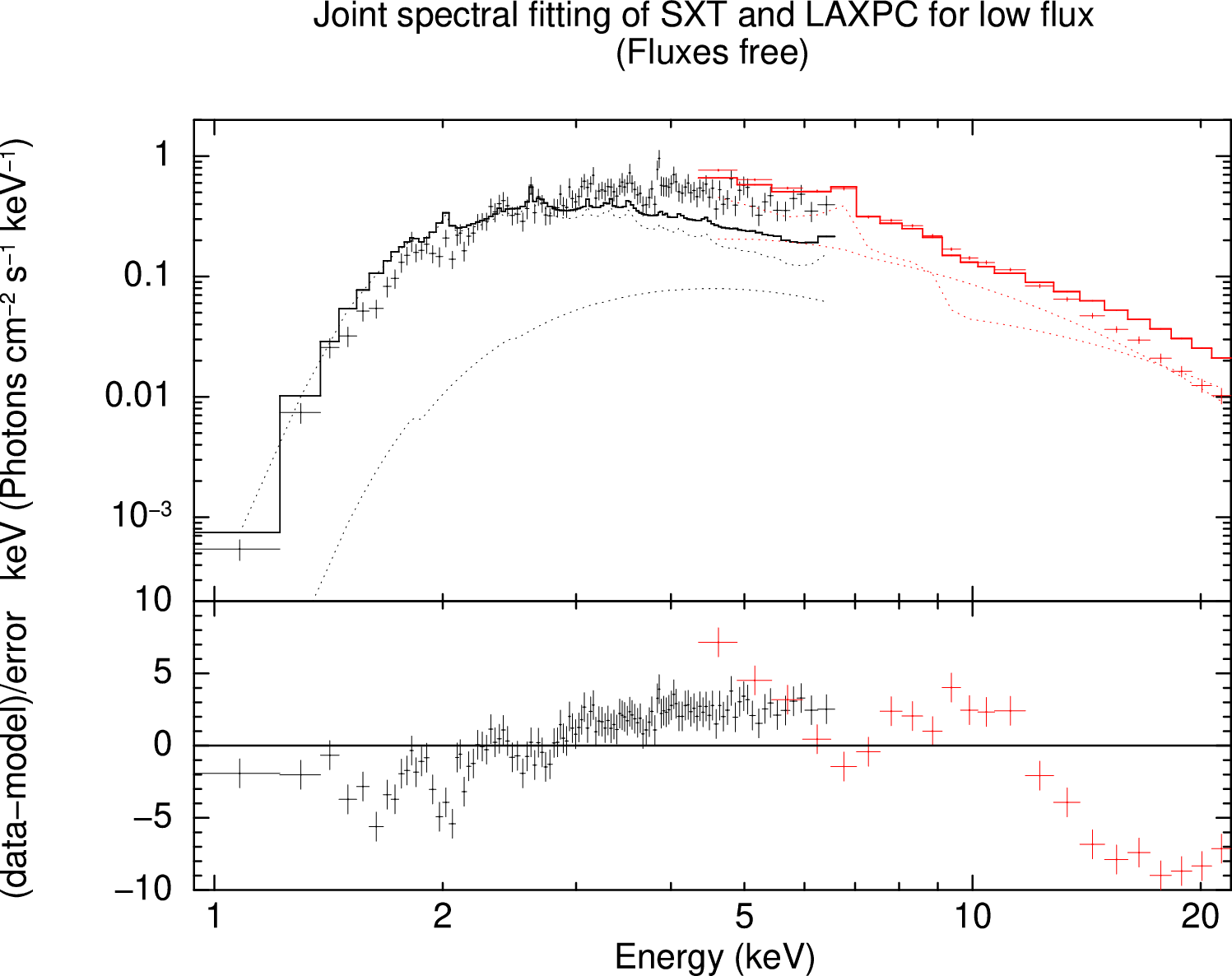}
    \hfill
    \includegraphics[width=0.37\textwidth,angle=270]{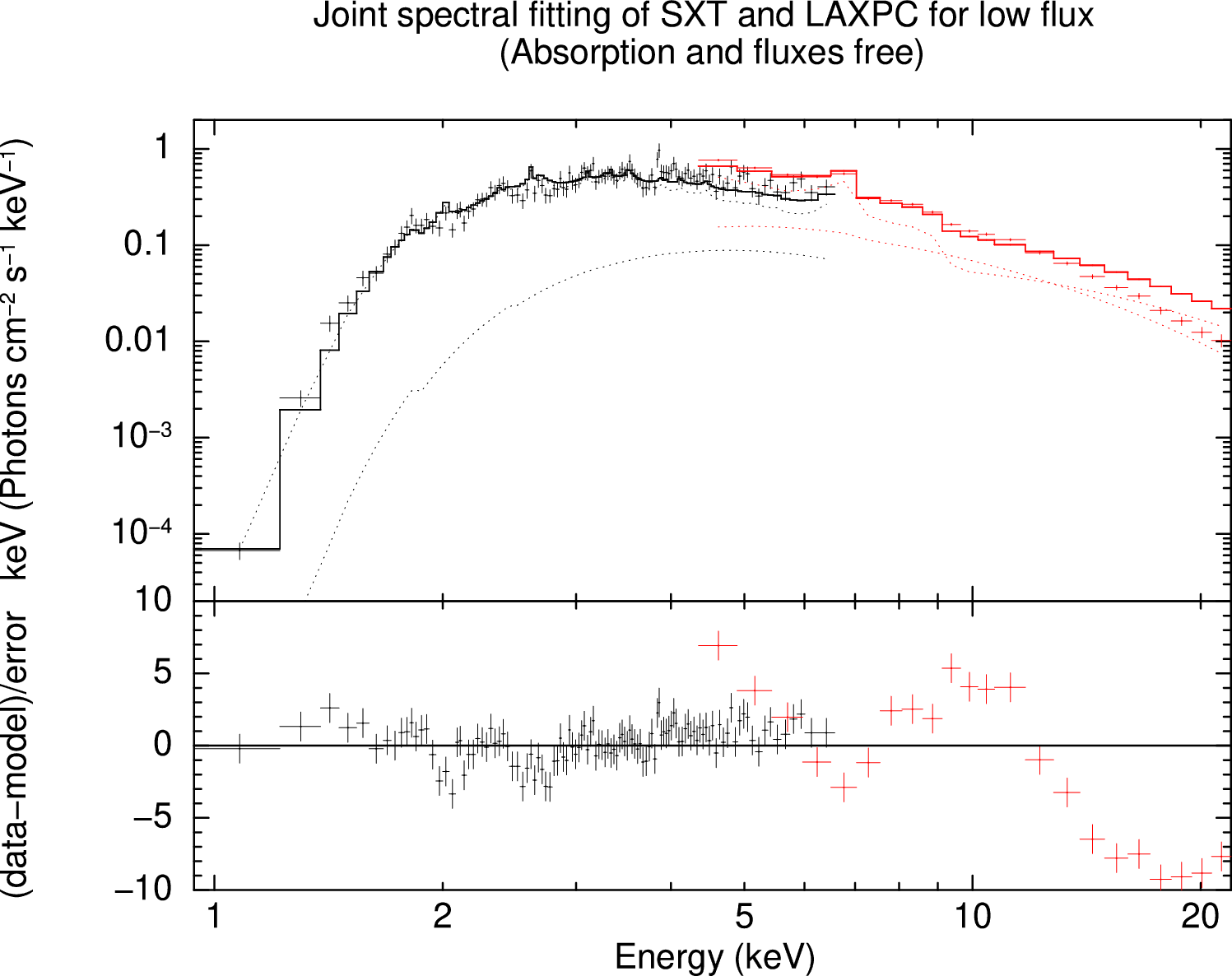}
    \hfill
    \includegraphics[width=0.37\textwidth,angle=270]{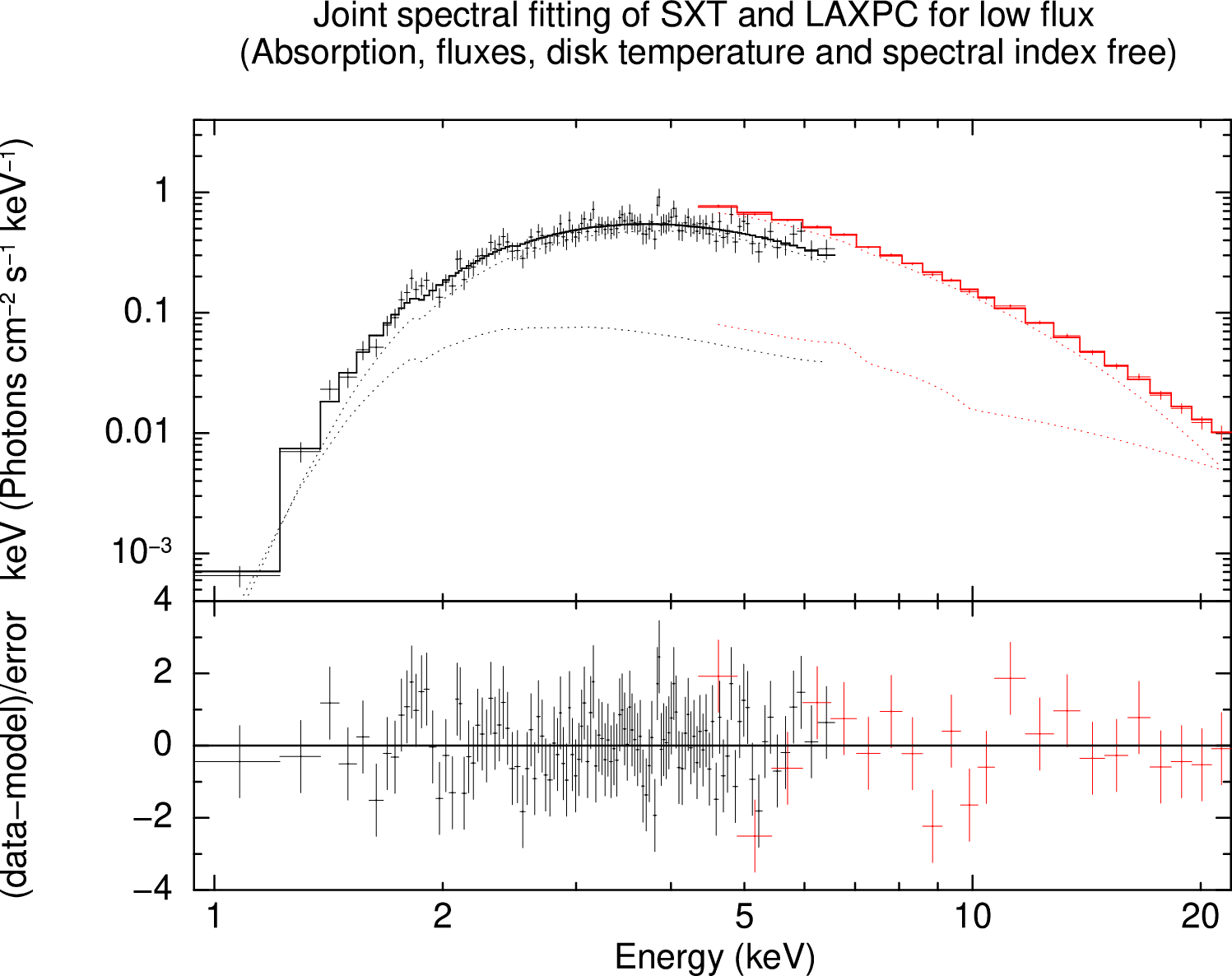}
    
    \caption{Unfolded \asat{} SXT/LAXPC20 mean spectra of \src{} in the 1.0–25.0 keV energy range. The upper-left panel shows the best-fit joint spectra for the high-flux state of the $\omega$ class (Obs. $\omega_2$), including the model components and data-to-model ratio. The upper-right and middle-left panels display the joint spectral fits for Case-A and Case-B, respectively, corresponding to the low-flux state of the $\omega$ class (Obs. $\omega_2$). The lower-right panel presents the best-fit joint spectra for the low-flux (Case-C) of the $\omega$ class.}
    \label{fig8}
\end{figure*}
\begin{figure*}[!ht]
    \centering
    \includegraphics[width=0.49\textwidth]{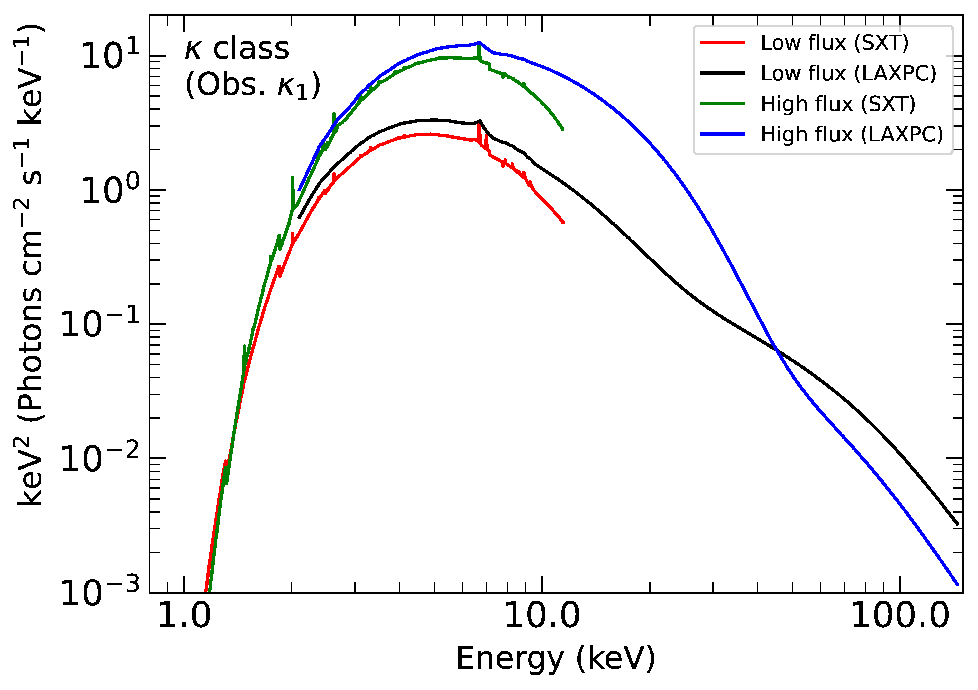}
    \hfill
    \includegraphics[width=0.49\textwidth]{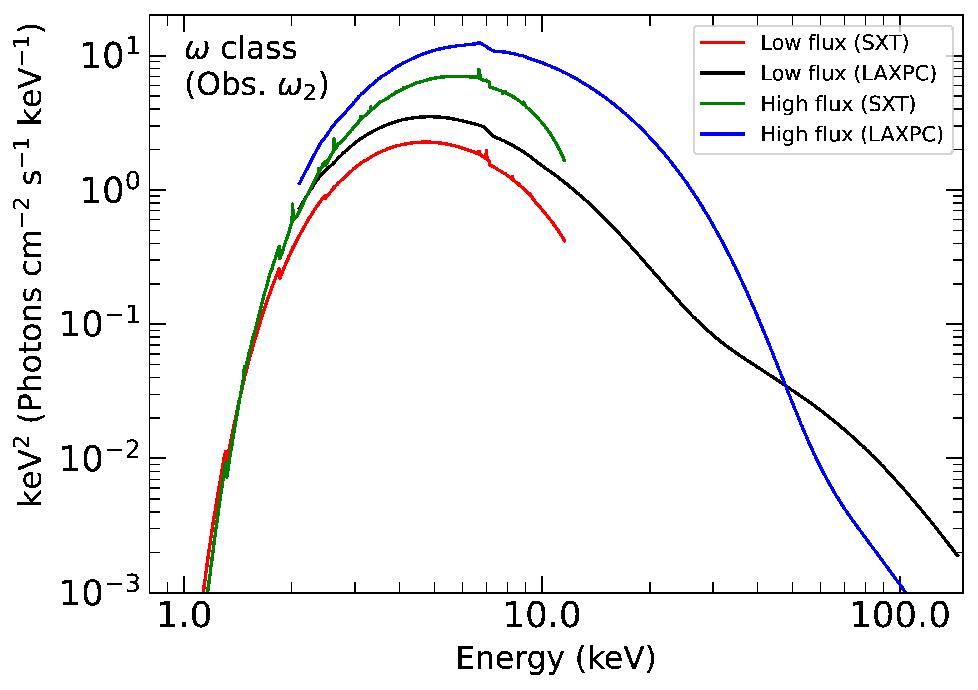}
    \caption{The left panel presents the best-fit continuum models for both the high and low-flux observations of the $\kappa$ class (Obs. $\kappa_1$),  while the right panel shows the best-fit continuum models for both the high and low-flux observations of the $\omega$ class (Obs. $\omega_2$). }
    \label{fig9}
\end{figure*}

\end{document}